\documentclass[12pt,preprint]{aastex}

\shorttitle{Gamma-Ray Burst Hubble Diagram}
\shortauthors{Schaefer}

\begin{document}

\title{The Hubble Diagram to Redshift $>$6 from 69 Gamma-Ray Bursts}

\author{Bradley E. Schaefer}
\affil{Physics and Astronomy, Louisiana State University,
    Baton Rouge, LA, 70803}

\begin{abstract}

The Hubble diagram (HD) is a plot of the measured distance modulus versus the measured redshift, with the slope giving the expansion history of our universe.  In the late 1990's, observations of supernova out to a redshift of near unity demonstrated that the universal expansion is now accelerating, and this was the first real evidence for the mysterious energy we now call Dark Energy.  One of the few ways to measure the properties of Dark Energy is to extend the HD to higher redshifts.  Many models have been proposed that make specific predictions as to the shape of the HD and so this offers a means of testing and eliminating models.  Taking the HD to high redshifts provides a way to test models where their differences are large.  For example, in a comparison of the now concordance model (a flat universe with $\Omega_M=0.27$ or so and constant Cosmological Constant) with a representative model of evolving Dark Energy (for which I'll take $w(z)=-1.31 + 1.48z$ from Riess' analysis of the gold sample of supernovae), the predicted distance moduli differ by 0.15 mag at $z=1.7$ and 1.00 mag at $z=6.6$.  The only way to extend the HD to high redshift is to use Gamma-Ray Bursts (GRBs).  GRBs have been found to be reasonably good standard candles in the usual sense that light curve and/or spectral properties are correlated to the luminosity, exactly as for Cepheids and supernovae, then simple measurements can be used to infer their luminosities and hence distances.  GRBs have at least five properties (their spectral lag, variability, spectral peak photon energy, time of the jet break, and the minimum rise time) which have correlations to the luminosity of varying quality.  All of these properties provide independent distance information and their derived distances should be combined as a weighted average to get the best value.  For GRBs which have an independently measured redshift from optical spectroscopy, we have enough information to plot the burst onto the HD.  In this paper, I construct a GRB HD with 69 GRBs over a redshift range of 0.17 to $>$6, with half the bursts having a redshift larger than 1.7.  This paper uses over 3.6 times as many GRBs and 12.7 times as many luminosity indicators as any previous GRB HD work.  For constructing the GRB HD, it is important to perform the calibration of the luminosity relations for every separate cosmology considered, so that we are really performing a simultaneous fit to the luminosity relations plus the cosmological model.  I have made detailed calculations of the gravitational lensing and Malmquist biases, including the effect of lensing de/magnification, volume effects, evolution of GRB number densities, the GRB luminosity function, and the discovery efficiencies as a function of brightness.  From this, I find that the biases are small, with an average of 0.03 mag and an RMS scatter of 0.14 mag in the distance modulus.  This surprising situation arises from two causes, the first being that burst peak fluxes above threshold do not vary with redshift and the second being that the four competing effects nearly cancel out for most GRBs.  The GRB HD is well-behaved and nicely delineates the shape of the HD.  The reduced chi-square for the fit to the concordance model is 1.05 and the RMS scatter about the concordance model is 0.65 mag.  This accuracy is just a factor of 2.0 times that gotten for the same measure from all the big supernova surveys.  I claim that GRBs will not suffer from effects due to evolution in the progenitors as we look back in time, with the reason being that the luminosity indicators are results of light travel time delays, conservation of energy in the shocked material, and the degree of relativistic beaming, with these not changing with metallicity or age.  That is, even though distant bursts might be more luminous on average than nearby bursts, the luminosity indicators will still operate to return the correct luminosity.  I fit the GRB HD to a variety of models, including where the Dark Energy has its equation of state parameter varying as $w(z)=w_0 + w_a z/(1+z)$.  I find that the concordance model is consistent with the data.  That is, the Dark Energy can be described well as a Cosmological Constant that does not change with time.

\end{abstract}



\section{Introduction}

	The Hubble diagram (HD) is a plot of distance versus redshift, with the slope giving the expansion history of our Universe.  This expansion history depends on the amount of mass in our Universe (both normal and dark matter) as well as the Dark Energy.  So by making observations of distances over a wide range of redshifts, we can measure the expansion history and place significant constraints on models of the Universe.  Indeed, it was this program that lead to the discovery of Dark Energy (Perlmutter et al. 1999), while observational knowledge about Dark Energy arises solely from the constraints on the expansion history.  The program of measuring the HD with supernovae has now occupied a vast amount of telescope time during the last decade and it will likely lead to a dedicated satellite over the next decade.
	
	Supernovae are now seen as near-ideal 'standard candles' for purposes of the Hubble diagram.  (The term 'standard candle' is commonly used to indicate a light source by which the luminosity can be determined regardless of whether the objects in that class all have the same luminosity.  For example, both Cepheids and Type Ia supernovae are called standard candles even though they vary in absolute magnitude by up to four magnitudes.)  But it is worthwhile remembering that only a decade ago the situation was that only one moderate-redshift event had ever been seen (N\o rgaard-Nielsen et al. 1989) and there was still widespread debate as to whether supernovae even were 'standard candles' (van den Bergh 1993; 1996; van den Bergh \& Pazder 1992; Phillips 1993; Tamman \& Sandage 1995).  This situation changed only when Hamuy et al. (1996) used a large and well-observed data set to prove that Type Ia supernovae were indeed 'standard candles', Perlmutter et al. (1995) found efficient methods to discover distant events, Kim, Goobar, \& Perlmutter (1996) solved the k-correction problem, and Perlmutter et al. (1997) started reporting on high-redshift light curves.  As is often the case with large and complex programs, the preliminary results (Perlmutter et al. 1997) gave a completely different conclusion than the more settled later conclusions (Perlmutter et al. 1998; 1999).  The nearly simultaneous results from an independent group of supernova hunters (Riess et al. 1998) provided the community with a good sense about the validity of the conclusions.  Later massive campaigns involving ground-based (Astier et al. 2006) and HST observations (Riess et al. 2004) have confirmed the earlier results.  Nevertheless, as is usual for important claims, a variety of troubles were raised, including grey-dust (Aguirre 1999a; 1999b), evolution of the supernova progenitors (Dom\'{i}nguez, H\"{o}flich, \& Straniero 2001), and refraction by Lyman-$\alpha$ clouds (Schild \& Dekker 2005).  But what really convinced the astronomy and physics community is that other independent lines of evidence have now been made that confirm the cosmological parameters derived from supernovae (Lineweaver 1999; Spergel et al. 2003; Eisenstein et al. 2006).  These results are consistent with a cosmology where the Universe is flat with $\Omega_M =0.27$ and $\Omega_{\Lambda} =0.73$, in which the Universe has an age of close to 14 billion years and will expand forever.  The observed acceleration of the universal expansion is attributed to a mysterious energy labeled 'Dark Energy'.
	
	The best way to measure properties of the Dark Energy seems to be to measure the expansion history of our Universe.  To this end, the SNAP satellite (Scholl et al. 2004) has been proposed to determine the distances and redshifts of two thousand supernovae per year out to redshift 1.7 with exquisite accuracy.  The default expectation is the simplest model for the Dark Energy, where it does not change in time.  This can be parameterized with the equation of state of the Dark Energy behaving as $p=w \rho c^2$, where $p$ is the pressure, $\rho$ is the density, $c$ is the speed of light, and $w$ is a dimensionless constant that might change with time.  The concordance case has $w=-1$ at all times, and this is the expectation of Einstein's cosmological constant, or if the Dark Energy arises from vacuum energy.  Given the strong results from supernovae for redshifts of less than 1, the frontier has now been pushed to asking the question of whether the value of $w$ changes with time (and redshift).  Lacking any physical understanding of such changes, we can parameterize the changes with low-order expansions such as $w(z)=w_0 + w\arcmin z$ or $w(z)=w_0 + w_a  z (1+z)^{-1}$ (Linder 2003).  So far the Supernova Legacy Survey results (Astier et al. 2006) have not been used to check the constancy of $w$, but the 'gold sample' of supernovae (Riess et al. 2004) shows that the best fit has $w_0 = -1.31$ and $w\arcmin = 1.48$.  
	
	If the Dark Energy changes with redshift, the best plan is to measure it over a wide range of redshifts, but supernovae cannot be detected out past redshift 1.7 even with the next generation of satellites.  Many alternatives to the concordance ($w=-1$) case have been proposed, including Weyl gravity (Mannheim 2006) and many versions of quintessence (Szyd\l owski, Kurek, \& Krawiec 2006), and at this point there is no strong reason to prefer the concordance.  The concordance cosmology makes a unique prediction as to the shape of the HD out to high redshift whereas alternatives make distinct and different predictions, so here is a perfect case where observations can test models of the Dark Energy.  After all, the concordance cosmology has only been around for a few years, whereas the previous concordance case (with $\Omega_M = 1$ and $\Omega_{\Lambda}=0$) was held just as dearly until its demise.  So a strong imperative in the quest for Dark Energy is to extend the HD to high redshift.
	
	Gamma-Ray Bursts (GRBs) offer a means to extend the Hubble diagram to at least redshifts of $>$6.  The reason is that GRBs are visible across much larger distances than supernovae.  For example, the average redshift of GRBs detected by the Swift spacecraft (Gehrels et al. 2004) is 2.8, while the range is from near zero out to $>$6 (Jakobsson et al. 2006).  GRBs are now known to have several light curve and spectral properties from which the luminosity of the burst can be calculated (once calibrated), and these make GRBs into 'standard candles'.  Just as with supernovae, the idea is to measure the luminosity indicators, deduce the source luminosity, measure the observed flux, and then use the inverse-square law to derive the luminosity distance for plotting on a HD.  To be placed on the HD, each GRB must have an independently measured redshift, usually from absorption lines in the optical spectra of the GRB afterglow or from optical spectra of the host galaxy.  To date, roughly 95 GRBs have a measured redshift (Greiner 2006).
	
	I presented the first GRB Hubble diagram in three public talks in March and April 2001 based on 8 GRBs (expanded to 9 GRBs in Schaefer 2001).  The first published GRB HD appeared in Schaefer (2003a) based on 9 GRBs using two luminosity indicators.  Half a year later, another GRB HD appeared with 16 bursts based on the independent result that the gamma-ray energy of the bursts is a constant after correction for the beaming angle (Bloom et al. 2003).  Since then, various groups have used one single luminosity indicator each to construct HDs or to similarly constrain cosmology (see Table 1).  This table shows that all prior work has been made with only a small fraction of the available data; both taking only a fraction of the useable bursts and in using only one or two luminosity indicators while more information from other indicators is ignored.  All the prior GRB HDs have had too large error bars to provide useful constraints on cosmology.
	
	The purpose of this paper is to make use of all possible GRBs and to make simultaneous use of all possible luminosity indicators.  As such, this paper will use over one order of magnitude more luminosity indicators than any previous work (see Table 1).  A preliminary version of this work was presented at the January 2006 meeting of the American Astronomical Society (Schaefer 2005), where I reported that the first results based on 52 GRBs showed a 2.5-sigma deviation from the unique prediction of the concordance cosmology.  Now, with 17 new GRBs plus substantial amounts of new data for the original 52 events as well as with some small improvements in analysis method (see below), the conclusions presented in this paper have a somewhat smaller deviation than in my preliminary results.  Just as with the change between the preliminary and final results of Perlmutter et al. (1997; 1999), I expect that the results in this paper will be closely checked by independent analysts, duplicated with independent events (collected over the next two years by Swift), and tested for potential problems.  A lesson learned from the supernova cosmology experience is that any results from the HD for $1.7< z \le 6.6$ will only be accepted by the community after independent methods reproduce the same conclusions.
	
\section{Observations}

	For a GRB to be placed on the HD, it must have a measured brightness, measured luminosity indicators, and an independent redshift.  The brightness and luminosity indicators require that a light curve and/or a spectral fit be available, while the redshift requires an optical spectrum of the afterglow or the host galaxy.  Greiner (2006) keeps an updated web page with all the bursts and which ones have reported redshifts, and this is the primary basis for selecting bursts.  This section will tell about the bursts selected and their measured properties.
	
	Out of the roughly 95 GRBs with reported redshifts, I have identified 69 to-date (with a cutoff at 7 June 2006) as having adequate data for placement on the GRB HD.  These are listed in Table 2.  
	
	Some bursts with reported redshifts are not included for a variety of reasons:  GRBs 050509, 050709, 050724, 051221, and 060502B are all short-hard bursts that belong to a different population than normal GRBs, and the luminosity indicators do not apply to this separate class of bursts.  GRB 051227 might be a short-hard burst and it only has a limit on its redshift.  GRBs 980329 and 011030 only have limits on their redshifts.  GRBs 980326, 011121, 020305, 050803, and 060123 have redshifts of too low a confidence to be used.  GRBs 050730, 050814, 051016, 060218, 060512, and 060522 are all too faint to have any useful constraints on their luminosity indicators.  GRBs 000214, 000418, 000301C, and 031203 do not have adequate light curve or spectral data available.  GRBs 980425, 020819, and 050315 are outliers of various types and will be handled separately in section 3.
	
	For each selected GRB, I have tabulated a range of primary information in Table 2.  Column 1 gives the six-digit GRB identification number in the usual format of YYMMDD.  Column 2 identifies the satellite experiment that is the source of the light curve used in each case.  Column 3 gives the redshift ($z$) of the GRB rounded to the nearest 0.01.  Column 4 gives the reference number (see the footnote to the table) for the source of the redshift.  Column 5 gives the observed peak flux ($P$) and its one-sigma uncertainty over the brightest one-second time interval during the burst.  Some uncertainties are not reported in the literature, so I have adopted a value of $P/10$ as the error bars for $P$ as indicated by values in square brackets.  The units are either photons/cm$^2$/s for values larger than 0.01 or in units of erg/cm$^2$/s for values smaller than 0.01 (as expressed in scientific notation).  All but four of the GRBs have the quoted $P$ values for a time interval length of 1 second (or 1.024 seconds), and the exceptions are marked with a footnote that also gives a conversion factor to convert to  a 1 second time interval as based on direct calculation from the observed light curves.  Column 6 gives the quoted energy range for the given $P$ values, $E_{min}$ and $E_{max}$.  Column 7 gives the reference for the quoted peak flux data.  Column 8 gives the peak energy of the $\nu F_{\nu}$ spectrum ($E_{peak}$).  The error bars are often asymmetric and so I first give the uncertainty in the direction of lower $E_{peak}$ followed by the uncertainty in the direction of higher $E_{peak}$.  Again, one-sigma error bars are often not quoted in the literature and so I have adopted the typical values as shown in square brackets.  For some GRBs, the $E_{peak}$ is not reported and so I have adopted an approximate $E_{peak}$ value based on the best fit luminosity relation for $E_{peak}$ along with very large adopted error bars.  It is important to note that these adopted $E_{peak}$ values are only used for extrapolating the spectra to the bolometric value and this assumption has little effect because the observations cover the energies with most of the flux.  Columns 9 and 10 give the values for the low-energy and high-energy power law spectral indices ($\alpha$ and $\beta$ respectively) along with their one-sigma uncertainties.  When a value of $\alpha$ is not known I adopt a value of $-1.1 \pm 0.4$.  When a value of $\beta$ is not known I adopt a value of $-2.2 \pm 0.4$. These spectral parameters are only used for estimating the bolometric peak fluxes and the adopted values make relatively little difference while the uncertainties are propagated into the bolometric peak fluxes.  Column 11 gives the reference for the spectral information.  
	
	Some GRBs have a measured jet break time ($t_{jet}$) when the afterglow brightness has a power law decline that suddenly steepens due to the slowing down of the jet until the relativistic beaming roughly equals the jet opening angle, $\theta _{jet}$ (Rhoads 1997).  The measured $t_{jet}$ value can be used to deduce $\theta _{jet}$ and hence convert the 'isotropic' energy ($E_{\gamma ,iso}$) into the total energy in gamma-rays emitted by the burst ($E_{\gamma}$).  From Sari, Piran, \& Halpern (1999), 
\begin{equation}
\theta_{jet} = 0.161 [t_{jet}/(1+z)]^{3/8} (n~\eta _{\gamma} / E_{\gamma ,iso,52})^{1/8}, 
\end{equation} 
where $z$ is the redshift, $t_{jet}$ is the jet break time measured in days, $n$ is the density of the circumburst medium in particles per cubic centimeter, $\eta _{\gamma}$ is the radiative efficiency, and $E_{\gamma ,iso,52}$ is the isotropic energy in units of $10^{52}$ erg for only an Earth-facing jet.  Here, $\theta_{jet}$ is in degrees and it is the angular radius (i.e., the half opening angle) subtended by the jet.  The isotropic energy is calculated as 
\begin{equation}
E_{\gamma ,iso} = 4\pi d_L^2 S_{bolo} (1+z)^{-1}, 
\end{equation} 
where $d_L$ is the luminosity distance of the burst and $S_{bolo}$ is the bolometric fluence of the gamma-rays in the burst.  The total collimation-corrected energy of the GRB is then
\begin{equation}
E_{\gamma} = (1-cos\theta _{jet})~E_{\gamma ,iso}, 
\end{equation} 
where the beaming factor, $F_{beam}$ is $1-cos\theta _{jet}$.

	One of the luminosity relations connects $E_{\gamma}$ with $E_{peak}$, so bursts with an observed jet break can be used as a luminosity indicator.  Unfortunately, only a little less than half of the bursts have observed jet breaks.  The important observational quantities have been listed in Table 3 for these bursts.  Column 1 gives the GRB name.  Column 2 lists the observed fluence ($S$) and its uncertainty.  Some error bars are not reported in the literature, so I have adopted a conservative 10\% error as indicated by square brackets.  All values are in units of erg/cm$^2$.  Column 3 gives the energy range (in units of keV) for the reported fluence, $E_{min}$ and $E_{max}$.  Column 4 gives a reference for the source of the quoted fluence.  Column 5 gives the bolometric fluence and its uncertainty (in units of erg cm$^{-2}$) as discussed in the next paragraph.  Column 6 lists the measured time of the jet break ($t_{jet}$) in units of days, while column 7 gives the reference for these values.  Columns 8 and 9 give the derived $\theta _{jet}$ values (in degrees) and the reference.  For many bursts, the value of $\theta _{jet}$ is based on a detailed model fitting of light curves in many energy bands (e.g., Panaitescu \& Kumar 2002).  In the absence of these detailed fits, it is reasonable to adopt $\eta _{\gamma}=0.2$ and $n=3~cm^{-3}$ as these values are present in the equation for $\theta_{jet}$ to the 1/8 power.  Finally, column 10 gives the derived value for $F_{beam}$ and its uncertainty.

	The reported brightnesses (peak fluxes and fluences) are given over a wide variety of observed band passes, and with the wide range of redshifts these correspond to an even wider range of energy bands in the rest frame of the GRB.  One way to put all these brightnesses onto a consistent basis is to derive the bolometric brightness, where the measured spectrum is extrapolated to high and low energies and integrated over all energies.  This can be done with fair accuracy for GRBs as the bulk of the energy comes out in and near the observed band passes so that uncertainties in the extrapolation are generally small.  In practice, I will integrate under the measured spectral shape from photon energies of 1 to 10,000 keV in the rest frame of the GRB.  I adopt the shape of the GRB spectrum to be a smoothly broken power law (Band et al. 1993) as
\begin{eqnarray}
\Phi (E) = \left\{ \begin{array}{ll}
	 A~E^{\alpha}~ e^{-(2+\alpha)E/E_{peak}} & \mbox{if $E \le [(\alpha - \beta)/(2+\alpha)]E_{peak}$} \\
	 B~E^{\beta} & \mbox{otherwise}
	 \end{array}
\right. \
\end{eqnarray} 
Here $\Phi$ is the usual differential photon spectrum ($dN/dE$) as a function of the photon energy ($E$), $E_{peak}$ is the photon energy at which the $\nu F_{\nu}$ spectrum is brightest, $\alpha$ is the asymptotic power law index for photon energies below the break, and $\beta$ is the power law index for photon energies above the break.  The two normalization parameters ($A$ and $B$) are chosen to ensure continuity at the break and to match the observed burst brightness.  If the observed peak flux ($P$) is reported in units of erg/cm$^2$/s, then the bolometric peak flux will be 
\begin{equation}
P_{bolo} = P  \int_{1/(1+z)}^{10^4/(1+z)} E \Phi dE ~~ / ~~ \int_{E_{min}}^{E_{max}} E \Phi dE 
\end{equation}
If the units of $P$ are photon/cm$^2$/s, then 
\begin{equation}
P_{bolo} = P  \int_{1/(1+z)}^{10^4/(1+z)} E \Phi dE ~~ / ~~ \int_{E_{min}}^{E_{max}}  \Phi dE 
\end{equation}
The result will be the bolometric peak flux over a one second interval with units of erg/cm$^2$/s.  Similarly, for the observed fluences ($S$ all in units of erg/cm$^2$), we can calculate the bolometric fluence as 
\begin{equation}
S_{bolo} = S  \int_{1/(1+z)}^{10^4/(1+z)} E \Phi dE ~~ / ~~ \int_{E_{min}}^{E_{max}} E \Phi dE .
\end{equation}
The uncertainties on both $P_{bolo}$ and $S_{bolo}$ are both calculated by propagating the uncertainties in $P$, $S$, $\alpha$, and $\beta$.  The resulting bolometric peak fluxes and fluences are tabulated in Table 4.  The bolometric fluences are given in Table 4 only for those bursts with known jet break times, as that is all that is needed for the purposes of this paper.  The isotropic luminosity is given by
\begin{equation}
L = 4\pi d_L^2 P_{bolo}. 
\end{equation} 
To calculate either $L$ or $E_{\gamma}$ (for use in the luminosity relations), we have to know the luminosity distance ($d_L$), which depends on the cosmological model and the redshift.

	The lag ($\tau_{lag}$) of a GRB is the time shift between the hard and soft light curves, with the soft photons coming somewhat later than the hard photons.  Schematically, we can view this as the delay in the time of peaks in the, say, 100-300 keV versus 25-50 keV energy bands.  But in practice, only the brightest bursts have their times of peaks defined well enough to make this a useful definition.  For most bursts, the normal Poisson noise in the light curve will create substantial errors if only the bins with the highest fluxes are compared.  A practical definition of the lag can be taken as the offset that produces the maximum cross correlation between the hard and soft light curves.  And for faint bursts, the cross correlation itself will have substantial noise so that it is best to fit a parabola around the maximum.  To avoid the noise added by time intervals when the light curve is dominated by ordinary background noise, the cross correlation is performed only for times when the total light curve (summed over all available energy bands) is brighter than 10\% of the peak flux in that light curve.  The choice of energy bands for the hard and soft light curves should be such that there is as wide a separation as possible yet with the high energy band still having significant flux.  A good choice is the original choice for the BATSE data (Band 1997; Norris, Marani, \& Bonnell 2000), with the soft band being from 25-50 keV and the hard band being from 100-300 keV.  The energy bands for the various experiments are often fixed, yet choices can be made that are reasonably close to this standard (see Table 5).  The individual lags calculated for each burst are given in Table 4.  Some GRBs (those from Konus and SAX) do not have two channel light curves available, while others are too faint and hence have the uncertainty in the lag too large to be useful.
	
	The variability ($V$) of a burst is a quantitative measure of whether its light curve is spiky or smooth.  A reasonable measure of this is to calculate the normalized variance of the observed light curve around a smoothed version of that light curve (Fenimore \& Ramirez-Ruiz 2000).  Unfortunately, this one-sentence description has a number of free parameters that can greatly change the calculated $V$.  Fenimore \& Ramirez-Ruiz (2000) and Reichart et al. (2001) experimented with a variety of choices.  Schaefer, Deng, and Band (2001) selected effective choices as being those that produced the best correlation between $V$ and $\tau_{lag}$ for 112 BATSE bursts, finding that the formulation of Reichart et al. was poor and that the final choices of Fenimore \& Ramirez-Ruiz were best.  As part of this paper, I have tried a wide variety of options in the definition of $V$  and have settled for the definition which yields the least scatter in the correlation between $V$ and the burst luminosity.  In this definition, I take the smoothed light curve ($C_{smooth}$ to be the original light curve ($C$) smoothed with box-smoothing where the width of the box is 30\% of the time for which the light curve is brighter than 10\% of its peak flux.  The counts per time bin in the background-subtracted light curve is $C$ with an uncertainty of $\sigma_C$, while the counts in the smoothed light curve are $C_{smooth}$.  With this, the variability is
\begin{equation}
V = \langle [(C-C_{smooth})^2-\sigma_C^2]/C_{smooth,max}^2 \rangle,
\end{equation}
where $C_{smooth,max}$ is the peak value of $C_{smooth}$, and the angle-brackets indicate an average over all time bins where the smoothed light curve is greater than 10\% of $C_{smooth,max}$.  For each satellite experiment, I have chosen a fairly broad energy band over which to construct the light curve (see Table 5).  With this definition, I have calculated $V$ for most of the GRBs and placed these values in Table 4.  Again, some GRBs are not included because the light curve might not be available, the light curve might have significant time gaps, or the burst brightness might be so faint that the derived uncertainty on $V$ was too large to be useful.

	The minimum rise time ($\tau_{RT}$) in the GRB light curve is taken to be the shortest time over which the light curve rises by half the peak flux of the pulse.  For bright bursts, it is an easy calculation to search time intervals before each peak for the shortest one in which the rise is half the peak brightness.  In cases where the rise from one time bin to the next is greater than half-peak, I take the rise time to be the appropriate fraction of the bin width.  But for faint bursts where the normal background noise provides large Poisson fluctuations, this procedure will always find fast rise times, so as a remedy I always bin the light curve until the uncertainty in individual points in the light curve are 10\% of the peak flux (or less).  In this binned light curve the fluctuations to mimic a rise by 50\% of the peak flux would have to be roughly 5-sigma, and this eliminates the effects of Poisson noise.  A cost of this binning is that rise times much shorter than the binned time intervals cannot be measured.  The energy bands used for these light curves are the same as used for calculating $V$ (see Table 5).  These rise times will depend on the exact choice of the first bin, so I have adopted the average of the derived rise times over all possible start bins as being the minimum rise time.  The uncertainty in the minimum rise time is then the RMS of the values for all the start bins.  These values are tabulated in Table 4.
	
	The number of peaks ($N_{peak}$) in a GRB light curve varies from one for a simple fast-rise-exponential-decay (FRED) light curve shape to a dozen or more for complex spiky bursts.  Let the overall maximum in the background subtracted light curve to be $C_{max}$.  I take a peak to be any local maximum which rises higher than $C_{max}/4$ and which is separated from all other peaks by a local minimum that is at least $C_{max}/4$ below the lower peak.  Again, this is simple to calculate for bright bursts but is not realistic for faint bursts with substantial Poisson noise.  This is solved by binning the GRB light curve such that each bin has a one-sigma uncertainty that is less than $C_{max}/12$.  As before, the $N_{peak}$ value can change with the phasing of the binning, but here the number of peak usually does not change.  The energy bands and time resolutions for the light curves are specified in Table 5.  The resultant $N_{peak}$ values are listed in Table 4.

	The luminosity relations will be power laws of either $L$ or $E_{\gamma}$ as a function of $\tau _{lag}$, $V$, $E_{peak}$, $\tau _{RT}$, and $N_{peak}$.  Both $L$ and $E_{\gamma}$ will have to be recalculated with luminosity distances appropriate for every cosmology.  Table 4 has compiled all the values needed for producing a GRB HD for any particular cosmology.  Columns 1 and 2 are the GRB identifier and the redshift.  Column 3 is $P_{bolo}$ (the flux over the peak 1 second interval from 1-10,000 keV in the rest frame of the GRB) and its uncertainty in units of erg/cm$^2$/s.  Column 4 gives $S_{bolo}$ and its uncertainty in units of erg/cm$^2$.  Column 5 lists the beaming factor, $F_{beam}$.  Column 6 tabulates the lag time for each burst in units of seconds in the Earth rest frame.  Column 7 gives the $V$ value.  Column 8 lists the observed $E_{peak}$ value from Table 2, provided that a definite measure of $E_{peak}$ is known.  Columns 9 and 10 present the values for $\tau_{RT}$ in seconds in the Earth rest frame and $N_{peak}$.

\section{Outliers}
		
	In any large set of observations taken from widely heterogeneous sources, there will be outliers.  Many of these outliers arise because of some sort of error, rather than simply being a tail of some ordinary distribution of measurement errors.  If these errors are allowed to remain, then they will dominate the fits and bias the results.  So the problem is to recognize the outliers without throwing out the error-free values.  I will adopt the standard solution of throwing out luminosity indicators or GRBs only if they deviate from best fit (with the outlier included) by 3-sigma or more.  For some of the outliers, I have rejected the use of the entire burst, while for others I will impeach only a specific measurement.  Here is a detailed list of the outliers that I have rejected:
	
	GRB 980425 is the smooth and faint burst which was identified with the supernova 1998bw in a nearby galaxy at z=0.0085.  From measures for the lag, $E_{peak}$, and the rise time, I get a combined distance modulus of 40.35 mag whereas the real distance modulus (independent of any choices for the cosmology) is 32.8 mag or so.  As with previous workers, I find that GRB 980425 is a distant outlier.  The reason is likely that this very low luminosity event is somehow greatly different than classical GRBs and so the luminosity indicators do not apply.

	GRB 990123 was the original optical transient observed {\it during} the burst by ROTSE (Akerlof et al. 1999).  Its redshift is a confident 1.61, and this is in good agreement with all five luminosity indicators {\it except} for the rise time.  The minimum rise time in this well measured light curve (2.3 s) is much too long for a burst of this high luminosity.  In the calibration curve for the $\tau_{lag}-L$ relation, GRB 090123 has its rise time as a distant outlier and this rise time is rejected.  I can only suggest that some bursts will happen to have their collisions all at far past the minimum radius and hence will have a long rise time for their luminosity.
	
	GRB 020819 was optically dark yet had a long fading radio transient that  provides an accurate position (Jakobsson et al. 2005b).  The indicated position is on top of a "blob" which appears near but definitely outside a barred spiral galaxy measured to be at $z=0.41$.  No redshift for the "blob" has been measured.  GRB 020819 is an outlier on the $\tau_{lag}$, $V$, $\tau_{RT}$, and $N_{peak}$ calibration curves.  The combined $\mu$ value is 44.27 mag while the barred spiral galaxy has a distance modulus of 41.74 mag.  Based on the luminosity indicators, the GRB is likely at $z > 1$ with a value around 1.7 being preferred.  I suggest that the "blob" is the real host galaxy at $z \sim 1.7$, and this prediction can be tested with a spectrum of the "blob".
	
	GRB 030328 has the same situation as GRB 990123, in that all five luminosity indicators are in good agreement with the observed redshift {\it except} for the rise time ($3.9 \pm 0.2$ s) despite the burst being bright enough that faster rises would be obvious in the light curve.  The two GRBs even have similar light curves.  Again, I will reject the rise time as being a distant  outlier.
	
	GRB 031203 does not have an available light curve or an $E_{peak}$ value, so it cannot be included in my analysis.  Nevertheless, this burst is an outlier for both the $\tau_{lag}-L$ and $E_{peak}-L$ relations (Sazanov, Lutovinov, \& Sunyaev 2004).  The trouble is that the reported measure of $\tau_{lag}=0.24 \pm 0.12 s$ and $E_{peak} > 190$ keV are both signs of a high luminosity event, whereas the burst is low luminosity at $z=0.105$.  It cannot be that the redshift is greatly in error since an underlying supernova (SN 2003 ) has been measured spectroscopically (Tagliaferri et al. 2004).  The general thinking is that GRB 031203 is another unusual low-luminosity event like GRB 980425 (Sazanov, Lutovinov, \& Sunyaev 2004; Soderberg et al. 2004; Watson et al. 2004) for which the luminosity relations do not apply.
	
	GRB 050315 is a far outlier for the $E_{peak}-L$ and $E_{peak}-E_{\gamma}$ relations, while it is on the edge for the $\tau_{RT}-L$ and on the best fit calibration line for the $V-L$ relation.  The combined $\mu$ is 43.71 mag while the distance modulus of $z=1.948$ is 45.87 mag.  The problem is unlikely to be an error in redshift.  This is because the outliers must be resolved by shifting to lower redshift yet the measured $z$ comes from absorption lines in the afterglow (Kelson \& Berger 2005) and thus it cannot be lowered.  A real problem for the $E_{peak}-E_{\gamma}$ relation is that the x-ray afterglow light curve has at least four breaks in it (Vaughan et al. 2006) so any assignment of $t_{jet}$ can only be problematic.  Another big problem for both the $E_{peak}-L$ and $E_{peak}-E_{\gamma}$ relations is that Vaughan et al. (2006) reports the $E_{peak}$ value ($16_{-7}^{+18}$ keV with error bars at the 68\% confidence level) to be at the edge of the spectral energy range for Swift (15-150 keV).  In this case, the detection of a break can only be problematic.  Both the $E_{peak}-L$ and $E_{peak}-E_{\gamma}$ relations would be easily satisfied if $E_{peak}$ is from 80--150 keV at the other edge of the Swift spectrum.  So my suggested solution is simply that $E_{peak}$ is really at $\sim 150$ keV where Swift cannot see it.  Despite the likelihood of this solution, GRB 050315 must remain a rejected outlier for my analysis.
	
	In all, I have rejected only three bursts as outliers plus two rise times as outliers.  This is fairly good given the many bursts and observations going into Tables 2-4.

\section{Luminosity Relations}
		
	The luminosity relations are connections between measurable parameters of the light curves and/or spectra with the GRB luminosity.  Specifically, I will be using the power law relationships between $\tau_{lag}-L$, $V-L$, $E_{peak}-L$, $E_{peak}-E_{\gamma}$, $\tau_{RT}-L$, and $N_{peak}-L$.  This section will discuss the calibration of all six relations.
	
	The calibration will essentially be a fit on a log-log plot of the luminosity indicator versus the luminosity.  For this calibration process, the burst's luminosity distance must be known to convert $P_{bolo}$ to $L$ (or $S_{bolo}$ to $E_{\gamma}$) and this is known only for bursts with measured redshifts.  However, an important point is that the conversion from the observed redshift to a luminosity distance requires some adopted cosmology.  This means that every cosmology will have a separate calibration.  Fortunately, as we will see below, the calibration only weakly depends on the input cosmology.  If we are interested in calibration for purposes of GRB physics, then it will be fine to adopt the calibration from some fiducial cosmology (say, the concordance cosmology with $w=-1$).  But if we are interested in testing the cosmology, then we have to use the calibration for each individual cosmology being tested.  For a particular cosmology, the theoretical shape of a HD has to be compared with the observed shape when the burst distances are calculated based on calibrations for that particular cosmology.  In practice, this means that the model and observed luminosity distances are compared in a chi-square sense as cosmological parameters vary with the observed values changing with the cosmology.  Thus, any test of cosmological models with a GRB HD will be a simultaneous fit of the parameters in the calibration curves and the cosmology.
	
	A comparison of this case with that of the supernova HD is instructive.  Schematically, the calibration of the luminosity relation for supernovae (i.e., decline rate versus luminosity) can be accomplished for nearby events for which the luminosity distances have no dependance on the cosmological parameters.  Once this calibration is learned from nearby events, then it can be applied to distant supernovae with confidence.  In this case, the calibration does not depend on cosmology and the deduced positions of the high redshift supernovae in a Hubble diagram will not depend on the cosmology being tested.  But in practice, many of the supernovae in Hamuy et al. (1996) are sufficiently distant that small effects of varying cosmology must be introduced, and then even the supernova HD becomes a simultaneous fit involving both calibrations and cosmologies.  Another feature of the supernova HD case is that it involves both nearby events where the distances come from Cepheids that are independent of the Hubble constant ($H_0$) and far events where the distances depend on the Hubble constant.  In such a case, distortions can appear in the shape of the HD depending on the adopted value for the Hubble constant.  To solve this problem, Perlmutter et al. (1997; 1999) adopted a method where the "Hubble-constant-free luminosity distance" is used along with the calibration with a "Hubble-constant-free B-band absolute magnitude at maximum", where they subtract $5\log H_0$ from the peak absolute magnitude ($M$) and add it back in again into the distance modulus ($\mu$) in the usual equation for the observed magnitude ($m$).  Thus, $m=M+\mu$ becomes $m=(M-5\log H_0)+(\mu+5\log H_0)$.  But with GRBs, we do not have the problem of mixing distances that are both dependent and independent of the Hubble constant.  So for GRBs, the "Hubble-constant-free" equation gives identical results as the normal equation.
	
	The conversion from redshift ($z$) to luminosity distance ($d_L$) depends on the cosmology.  Throughout this paper, I will assume that the Universe is flat, as indicated by strong sets of theoretical and observational arguments.  For a flat Universe with the concordance cosmology of $w=-1$, this relation is
\begin{equation}
d_L = cH_0^{-1} (1+z) \int_{0}^{z} dz' [(1+z')^3 \Omega_M + \Omega_{\Lambda}]^{-1/2}. 
\end{equation} 
Here, $\Omega_M$ is the dimensionless matter density and $\Omega_{\Lambda}=1-\Omega_M$ for a flat Universe.  If we allow for the possibility that the Dark Energy changes with time (i.e., $w$ varies with redshift), then we get a different relation.  In the absence of any real understanding of how Dark Energy changes (c.f. Szyd\l owski, Kurek, \& Krawiec 2006), a reasonable approach for analysis of observations is to simply expand $w$ as some function of redshift.  Riess et al. (2004) introduced the expansion $w(z)=w_0 + w\arcmin z$, but this has been widely criticized as being unsuitable at high redshifts because an exponential term grows large for $z>1$.  A better expansion (i.e., one that is well bounded at high redshift) is to adopt $w(z)=w_0 + w_a  z (1+z)^{-1}$ (Linder 2003).  This gives
\begin{equation}
d_L = cH_0^{-1} (1+z) \int_{0}^{z} dz' [(1+z')^3 \Omega_M + \Omega_{\Lambda}\{(1+z)^{3(1+w_0+w_a)}\}e^{-3w_az/(1+z)}]^{-1/2}. 
\end{equation} 
All such expansions are attempts to parameterize the redshift dependance of $w$ with a minimum of adjustable parameters.
	
	In addressing the question of testing whether the Dark Energy changes with redshift, it will be convenient to compare two cases: one representing an unchanging $w=-1$, while the other is representative of some reasonable case of a variable $w$.  Such a comparison can be helpful for seeing how much things change (e.g., the relative positions of the observed GRBs in the HD) between the constant and variable cases.  The only 'representative' case for a variable $w$ is that of the best fit for the 'gold sample' of supernovae (Riess et al. 2004) with $w_0 = -1.31$ and $w\arcmin = 1.48$.  In this paper, I will only use the $w(z)=w_0 + w\arcmin z$ expansion for purposes of making this comparison.  For all other purposes (like constraining the possible variations in the Dark Energy in Section 7), I will adopt the $w(z)=w_0 + w_a  z (1+z)^{-1}$ expansion of Linder (2003).

	The observed luminosity indicators will have different values from those that would be observed in the rest frame of the GRB.  That is, the light curves and spectra seen by Earth-orbiting satellites suffer time-dilation and redshift.  The physical connection between the indicators and the luminosity is in the GRB rest frame, so we must take our observed indicators and correct them to the rest frame of the GRB.  For the two times ($\tau_{lag}$ and $\tau_{RT}$), the observed quantities must be divided by $1+z$ to correct for time dilation.  The observed $V$ value varies as the inverse of the time stretching, so our measured value must be multiplied by $1+z$ to correct to the GRB rest frame.  The observed $E_{peak}$ value must be multiplied by $1+z$ to correct for the redshift of the spectrum.  The number of peaks in the light curve is defined in such a way as to have no $z$ dependance.  The dilation and redshift effects on $\theta_{jet}$ and $E_{\gamma ,iso}$ have already been corrected in equations 1 and 2.  A possibly substantial problem for the $\tau_{lag}$, $V$, and $\tau_{RT}$ relations is that we are in practice limited to the available energy bands (c.f. Table 5) whereas these correspond to different energy bands in the GRB reference frame.  Ideally, we would want to measure these indicators in observed energy bands that correspond to some consistent band in the GRB frame.
	
	The luminosity relations are all expressed as power laws and thus should be a linear function of the logarithms of the corrected indicator versus the logarithms of the burst luminosity.  These are displayed in Figures 1-6 for the concordance cosmology ($w=-1$ and $\Omega=0.27$ in a flat universe).  The scatter of the data about the best fits is consistent with a Gaussian distribution in this log-space.
	
	The plots of the luminosity relations show significant error bars in both the horizontal and vertical axes.  We would get different best fit slopes depending on whether we calculate a standard ordinary linear regression where the residuals in the luminosities are minimized or a linear regression where the residuals in the luminosity indicators are minimized.  The measurement uncertainties in both the luminosities and their indicators is smaller than the observed scatter about the best fit lines.  This means that there must be some additional source of intrinsic scatter which is dominating over the simple measurement errors.  And this scatter appears to be independent of both luminosity and redshift.  In this case,  the conclusion is to perform the ordinary least squares without any weighting.  (The use of weighted least squares results in almost identical best fits.)  The best fit values (from ordinary least squares with no weighting) are given in many reference books (e.g., Press et al. 1992).  In the luminosity relations, the two variables are not directly causative (with both being determined by $\Gamma_{jet}$), so the bisector of the two ordinary least squares (Isobe et al. 1990) will be used.  To be specific with the lag-luminosity relation as an example, with $Y_i= \log L$ and $X_i=\log(\tau_{lag})-\log(1+z)$, we are seeking the best fit to $Y_i=a+bX_i$.  The simple averages of $X_i$ and $Y_i$ are denoted as $\bar{X}$ and $\bar{Y}$.  The intercept is $a=\bar{Y}-b\bar{X}$.  For $\bar{X} \neq 0$, the best estimate of $a$ will be sensitive to small errors in $b$ and the confidence region for $a$ and $b$ will be a long thin ellipse.  To avoid these problems, we can recast the equation to be fit as $Y_i=\bar{Y} - b(X_i-\bar{X})$.
	
	In the following sections, I will present each of the luminosity relations and their derived best fit equations.  Also, for each relation, I will sketch out the theoretical justification and the physics behind each relation.  (In fact, three of these relations were theoretically predicted before they were confirmed with data.)  Despite have a theoretical understanding, the relations are essentially empirical, with the tightness of the relation being more important than any physical model.  This case is also true for the calibration of Type Ia supernovae.

\subsection{Lag versus Luminosity}

	The GRB lag time was first identified as a luminosity indicator by Norris, Marani, \& Bonnell (2000) based on six BATSE GRBs with optical redshifts.  This relation has a closely inverse relation between luminosity and lag, with very long lag events being very low luminosity and very short lag events being very high luminosity.  This relation (and the $V-L$ relation) was confirmed by Schaefer, Deng, \& Band (2001) based on the predicted $\tau_{lag}-V$ relation appearing for 112 BATSE bursts with no measured redshifts.  Norris (2002) demonstrated that the long lag bursts (hence the lowest luminosity and generally the closest) show a concentration to the supergalactic plane.
	
	The physics of the lag-luminosity relation is apparently simple (Schaefer 2004), and indeed the relation should have been deduced a decade ago as a necessary consequence of the empirical/theoretical Liang-Kargatis relation (Liang \& Kargatis 1996; Crider et al. 1999; Ryde \& Svensson 2000; 2002).  The Liang-Kargatis relation is that the time derivative of $E_{peak}$ is proportional to luminosity, as established by empirical measures or for the expected case where the cooling of the shocked material is dominated by radiative cooling.  Schaefer (2004) demonstrated that this time derivative equals a constant divided by the lag, so that we deduce $L \propto \tau_{lag}^{-1}$.  In essence, the shocked material will cool off at a rate dictated by its luminosity.  The lag is related to the time for a pulse to cool somewhat; if the burst is highly luminous then the cooling time (and hence the lag) will be short, while if the burst has a low luminosity then the cooling time and the lag will be long.
	
	The calibration data is plotted in Figure 1 as $\log [\tau_{lag}/(1+z)]$ versus $L$.  This plot is based on an assumed concordance cosmology ($w=-1$), $\Omega_M = 0.27$ (and hence $\Omega_{\Lambda} = 0.73$), and equation 10.  The best fit linear regression line is also plotted.  The equation for this calibration line is
\begin{equation}
\log L = 52.26-1.01\log[\tau_{lag}(1+z)^{-1}/0.1s].
\end{equation} 
  I note that the slope in this relation is satisfactorily close to the theoretical value of -1.  The one-sigma uncertainties in the intercept ($a=52.26$) and slope ($b=-1.01$) are $\sigma_a =0.06$ and $\sigma_b=0.05$.

	What is the uncertainty on any $L$ value deduced from this relation?  There will be two components that must be added in quadrature.  The first is the normal measurement error obtained by propagating the uncertainty through the last equation.  The second will be the addition of some systematic error that accounts for the extra scatter observed in Figure 1.  I will label this systematic error as $\sigma_{lag,sys}$.  Then, we have
\begin{equation}
\sigma_{\log L}^2 = \sigma_a^2 + (\sigma_b \log[\tau_{lag}(1+z)^{-1}/0.1s])^2 + (0.4343 b \sigma_{lag} / \tau_{lag})^2 + \sigma_{lag,sys}^2.
\end{equation} 
The value of $\sigma_{lag,sys}$ can be estimated by finding the value such that a chi-square fit to the lag-luminosity calibration curve produces a reduced chi-square of unity.  With this, I find $\sigma_{lag,sys} = 0.39$.

	How much does this calibration depend on the cosmology?  To get an idea of the characteristic changes, we can compare the calibration for the concordance $w=-1$ cosmology (i.e., equation 10) with that derived for the changing dark energy case given as the best fit to the 'gold sample' of supernovae ($w_0 = -1.31$ and $w\arcmin = 1.48$) from Riess et al. (2004).  This best fit cosmology from Riess produces calibration intercept and slope of 52.18 and -0.96.  As such, we see that the variation in the calibration of the lag-luminosity relation only weakly depends on typical range of cosmological models.
	
\subsection{Variability versus Luminosity}

	The GRB variability ($V$) was first recognized as a luminosity indicator by Fenimore \& Ramirez-Ruiz (2000) based on seven BATSE GRBs with optical redshifts.  The validity of the $V-L$ relation was confirmed with additional bursts by Reichart et al. (2001) and by the existence of the predicted $\tau_{lag}-V$ relation seen with 112 BATSE bursts without redshifts (Schaefer, Deng, \& Band 2001) and as updated to 551 BATSE bursts without redshifts (Guidorzi 2005).  Lloyd-Ronning \& Ramirez-Ruiz (2002) also confirmed the existence of the $V-L$ relation by demonstrating the existence of the predicted $V-E_{peak}$ relation for 159 BATSE GRBs with no known redshift plus 8 BATSE GRBs with optical redshifts.  Fenimore \& Ramirez-Ruiz (2000), Reichart et al. (2001), Schaefer, Deng, \& Band (2001), Guidorzi et al. (2005), and Li \& Paczy\'{n}ski (2006) have presented a variety of definitions of $V$ which use various smoothing functions and parameters and choice of time intervals.

	The origin of the $V-L$ relation is based in the physics of the relativistically shocked jets.  Detailed explanations within this model have been provided by M\'{e}sz\'{a}ros et al. (2002) and Kobayashi, Ryde, \& MacFadyen (2002).  In general, both $V$ and $L$ are functions of the bulk Lorenz factor of the jet ($\Gamma_{jet}$), where the luminosity varies as a moderately high power of $\Gamma_{jet}$ and where the high values of $\Gamma_{jet}$ make for smaller regions of visible emission and hence the faster rise times and shorter pulse durations result in high variability.
	
	The calibration plot for the $V-L$ relation is given in Figure 2 along with the best fit line.  This best fit can be represented with the equation
\begin{equation}
\log L = 52.49+1.77\log[V(1+z)/0.02].
\end{equation} 
The one-sigma uncertainties in the intercept and slope are $\sigma_a =0.22$ and $\sigma_b=0.12$.  The uncertainty in the log of the luminosity is 
\begin{equation}
\sigma_{\log L}^2 = \sigma_a^2 + (\sigma_b \log[V(1+z)/0.02])^2 + (0.4343 b \sigma_{V} / V)^2 + \sigma_{V,sys}^2.
\end{equation} 
The chi-square of the points about the best fit line is unity when $\sigma_{V,sys}=0.40$.

The best fit cosmology from Riess et al. (2004) produces calibration intercept and slope of 52.22 and 1.67 respectively.  This alternate cosmology changes the calibration parameters by about one-sigma.

\subsection{$E_{peak}$ versus Luminosity}

	$E_{peak}$ has been strongly correlated with both $L$ (Schaefer 2003b) and $E_{\gamma,iso}$ (Amati et al. 2002).  The two relations are likely caused by different physics, with the $E_{peak}-E_{\gamma,iso}$ relation being an approximation of the $E_{peak}-E_{\gamma}$ relation (discussed in the next subsection) that is related to the total energetics of the burst.  The $E_{peak}-L$ relation is different because it is related to the instantaneous physics at the time of the peak.  The idea is that the peak luminosity varies as some power of $\Gamma_{jet}$ while the $E_{peak}$ also varies as some other power of $\Gamma_{jet}$, so that $E_{peak}$ and $L$ will be correlated to each other through their dependance on $\Gamma_{jet}$.  Indeed, a detailed analysis of the situation resulted in a {\it prediction} that $L \propto [E_{peak}(1+z)]^{N/(M+1)}$ with $N/(M+1) \sim 2.5$, and this prediction was confirmed (Schaefer 2003b) from sets of 20 and 84 bursts.  This analysis also explains why the {\it observed} distribution of $E_{peak}$ can possibly be so narrow (Mallozzi et al. 1995) despite very wide ranges in both $\Gamma_{jet}$ and $1+z$, as well as explains why the observed average $E_{peak}$ varies as a particular function of the observed peak flux (Mallozzi et al. 1995).
	
	Figure 3 plots the values of $\log[E_{peak}(1+z)]$ versus $\log L$ for all bursts with available data.  The best fit line is plotted and is represented by the equation
\begin{equation}
\log L = 52.21+1.68\log[E_{peak}(1+z)/300keV].
\end{equation} 
The one-sigma uncertainties in the intercept and slope are $\sigma_a =0.13$ and $\sigma_b=0.05$.  The uncertainty in the log of the luminosity is 
\begin{equation}
\sigma_{\log L}^2 = \sigma_a^2 + (\sigma_b \log[E_{peak}(1+z)/300keV])^2 + (0.4343 b \sigma_{E_{peak}} / E_{peak})^2 + \sigma_{E_{peak},sys}^2.
\end{equation} 
The chi-square of the points about the best fit line is unity when $\sigma_{E_{peak},sys}=0.36$.

With Riess' best fit cosmology ($w_0 = -1.31$ and $w\arcmin = 1.48$), the intercept and slope are 52.11 and 1.60 respectively.  Again, the calibration is only weakly dependent on the input cosmology.

\subsection{$E_{peak}$ versus $E_{\gamma}$}

	Ghirlanda, Ghisellini, \& Lazzati (2004) discovered a tight correlation between $E_{peak}$ and $E_{\gamma}$.  This is an improvement on (and combination of) both the $E_{\gamma}=constant$ relation of Bloom et al. (2003) and the $E_{peak}-E_{\gamma ,iso}$ relation of Amati et al. (2002).  The physics of the $E_{peak}$ and $E_{\gamma}$ relation is well explained as a simple consequence of viewing geometry and relativistic effects within a standard jet model (Eichler \& Levinson 2004; Yamazaki, Ioka, \& Nakamura 2004; Rees \& M\'{e}sz\'{a}ros 2005; Levinson \& Eichler 2005).  This relation has the advantage of being one of the tightest for GRBs.  But this relation can only be used for the minority of GRBs with redshifts because a jet break has to be identified and measured in the afterglow light curve amongst the many bumps and breaks that are ubiquitous in afterglows.

	Figure 4 shows the calibration curve for the $E_{peak}-E_{\gamma}$ relation.  The best fit is 
\begin{equation}
\log E_{\gamma} = 50.57+1.63\log[E_{peak}(1+z)/300keV].
\end{equation} 
The one-sigma uncertainties in the intercept and slope are $\sigma_a =0.09$ and $\sigma_b=0.03$.  The uncertainty in the log of the burst energy is 
\begin{equation}
\sigma_{\log E_{\gamma}}^2 = \sigma_a^2 + (\sigma_b \log[E_{peak}(1+z)/300keV])^2 + (0.4343 b \sigma_{E_{peak}} / E_{peak})^2 + \sigma_{E_{\gamma },sys}^2.
\end{equation} 
The chi-square of the points about the best fit line is unity when $\sigma_{E_{\gamma},sys}=0.16$.

 With the Riess best fit cosmology, the intercept and slope are 50.50 and 1.59 respectively.

\subsection{$\tau_{RT}$ versus Luminosity}

	The variability of a light curve is a peculiar construction as we have no clear idea of what we are trying to measure and it is difficult to understand the physics of 'variability'.  In an effort to understand the meaning of variability, I calculated variability for a wide range of simulated light curves constructed from individual pulses with shapes as given by the model in Norris et al. (1996).  The most important determinant of the $V$ value was the rise time in the light curves, with other properties (like fall time, pulse duration, burst duration) being of lesser importance.  Thus, it seems that the $V$ might be just a surrogate measure for the rise time in the light curve.  And the rise time can be directly connected to the physics of the shocked jet.  Indeed, for a sudden collision of a material within a jet (with the shock creating an individual pulse in the GRB light curve), the rise time will be determined as the maximum delay between the arrival time of photons from the center of the visible region versus the arrival time of photons from the edge of the visible region.  Because the jet material is traveling at very close to light speed, this delay time is simply from the longer path length traveled (much like an echo).  This delay depends on the angular size (as viewed from the center of the GRB) of the visible region, which then depends on $\Gamma_{jet}$, so that rise times during a burst are proportional to $\Gamma_{jet}^{-2}$.  The proportionality constant depends on the radius from the GRB that the material is shocked.  But there is some minimum radius under which material is optically thick and inefficient at radiating and hence faint.  This minimum radius is roughly a constant from burst-to-burst (Panaitescu \& Kumar 2002), and this means that the minimum rise time should be roughly proportional to $\Gamma_{jet}^{-2}$.  The scatter in this relation will depend on how close the collisions in the jet occur to the minimum radius, so we could expect a substantial amount of scatter.  The burst luminosity also scales as $\Gamma_{jet}^N$ for $N \sim 3$.  With both $\tau_{RT}$ and $L$ being functions of $\Gamma_{jet}$, I {\it predicted} that the minimum rise time should be a luminosity indicator with $L \propto  \tau_{RT}^{-N/2}$ (Schaefer 2002).
	
	This prediction has proven true, for example as shown in Figure 5.  The best fit is
\begin{equation}
\log L = 52.54-1.21\log[\tau_{RT}(1+z)^{-1}/0.1s].
\end{equation} 
The one-sigma uncertainties in the intercept and slope are $\sigma_a =0.06$ and $\sigma_b=0.06$.  The uncertainty in the log of the burst energy is 
\begin{equation}
\sigma_{\log L}^2 = \sigma_a^2 + (\sigma_b \log[\tau_{RT}(1+z)^{-1}/0.1s])^2 + (0.4343 b \sigma_{\tau_{RT}} / \tau_{RT})^2 + \sigma_{RT,sys}^2.
\end{equation} 
The chi-square of the points about the best fit line is unity when $\sigma_{RT,sys}=0.47$.  

With the Riess cosmology, the intercept and slope are 52.42 and -1.14 respectively.

\subsection{$N_{peak}$ versus Luminosity}

	The number of peaks in a light curve depends on how many collisions between packets of material in the jet occur during the duration of the burst.  This number will be determined by many factors, including the exact realization of turbulence in the source and the distribution of velocities and densities in the jet.  However, some of the individual peaks might occur sufficiently close in time that these peaks will appear as one.  If the individual pulse durations are somewhat longer than the separation in time, then the two pulses will not be distinguishable as being separate.  The pulse durations ($D_{pulse}$) scale as the rise times (Nemiroff 2000) and hence will scale as $\Gamma_{jet}^{-2}$ or $L^{-2/N}$ (see previous subsection).  Schaefer (2003b) has presented several theoretical and observational arguments with the combined result that $N=3.14 \pm 0.34$.    For high luminosity bursts all collisions will result in distinct pulses in the light curve, while low luminosity events will have many of the collisions resulting in overlapping broad pulses.  Thus, a burst with many peaks can only be a high luminosity event because this is the only way to get narrow peaks that avoid merging together.  A burst with one or a few peaks could either be high luminosity (with few shell collisions) or low luminosity (with all the collisions producing merged peaks).  The maximum number of distinct peaks in a light curve with a duration of $T_{90}$ is $1+(T_{90}/D_{pulse})e^{-(1+D_{pulse}/T_{90})}$ for Poisson distribution of collisions.    With $D_{pulse}/T_{90} \simeq (L/L_0)^{-2/N}$ and $N \sim 3.14$, we can translate the observed number of peaks in a burst ($N_{peak}$) into a lower limit on the burst luminosity.  Again, this analysis was a theoretical {\it prediction} that was tested and shown to be true (Schaefer 2002).
	
	Figure 6 has a plot of the burst luminosity versus the number of peaks in a burst's light curve.  The curved line is the theoretical limit, and this was drawn for $\log L_0 = 50.08$ so as to tuck the limit up against the lower envelope of the data.  The important point is that there is an observational lower limit that corresponds in shape to the theoretical prediction, and this confirms the general picture.
	
	The theoretical limit is actually quite straight when plotted in log-log space.  So a convenient representation would be a simple power law
\begin{equation}
\log L \ge 50.32+2\log[N_{peak}]~~~~{\rm for}~ N_{peak} \ge 2.
\end{equation} 
For $N_{peak}=1$, there is no lower limit on the luminosity.  This approximation is plotted in Figure 6 as the thin straight line segment.

	As a limit, equation 22 has little utility for Hubble diagram purposes.  This means that I will only be using the first five luminosity indicators (sections 4.1-4.5) for the rest of this paper.  However, there is utility in the $N_{peak}-L$ relation for two other purposes.  First, it provides a startling theoretical prediction and observational correlation that should be studied for further insights into burst physics.  Second, counting the number of peaks is a fast way to spot some high luminosity GRBs.  For example, if Swift sees a faint burst with many peaks, then the burst must be at high redshift.
	
\subsection{Combining the Derived Distance Moduli}

	For every $L$ or $E_{\gamma}$ that we calculate, we can also derive a luminosity distance from the inverse-square law.  The equations for this are
\begin{equation}
d_L = [L/(4\pi P_{bolo})]^{1/2},
\end{equation} 
\begin{equation}
d_L = [E_{\gamma}(1+z)/(4\pi F_{beam} S_{bolo})]^{1/2}.
\end{equation} 
A distance modulus ($\mu$) can be calculated for every estimated luminosity distance as
\begin{equation}
\mu = 5 \log (d_L) -5,
\end{equation} 
with $d_L$ expressed in units of parsecs.  The propagated uncertainties will depend on whether $P_{bolo}$ or $S_{bolo}$ is used;
\begin{equation}
\sigma_{\mu} = [(2.5 \sigma_{\log L})^2+(1.086 \sigma_{P_{bolo}}/P_{bolo})^2]^{1/2},
\end{equation} 
or
\begin{equation}
\sigma_{\mu} = [(2.5 \sigma_{\log E_{\gamma}})^2+(1.086 \sigma_{S_{bolo}}/S_{bolo})^2+(1.086 \sigma_{F_{beam}}/F_{beam})^2]^{1/2}.
\end{equation} 
With five luminosity indicators, each burst will have up to five measured distance moduli and their one-sigma uncertainties, which I will label as $\mu_1 \pm \sigma_{\mu_1}$, $\mu_2 \pm \sigma_{\mu_2}$, $\mu_3 \pm \sigma_{\mu_3}$, $\mu_4 \pm \sigma_{\mu_4}$, and $\mu_5 \pm \sigma_{\mu_5}$ for the five indicators from subsections 4.1-4.5 respectively. 
	
	The best estimate $\mu$ for each GRB will be the weighted average of all available distance moduli.  Thus, the derived distance modulus for each burst will be 
\begin{equation}
\mu = [\sum_i \mu_i / \sigma_{\mu_i}^2]/[\sum_i \sigma_{\mu_i}^{-2}],
\end{equation} 
and its uncertainty will be
\begin{equation}
\sigma_{\mu} = [\sum_i \sigma_{\mu_i}^{-2}]^{-1/2},
\end{equation} 
where the summations run from 1--5 over the indicators with available data.  The weighted average formalism takes care of the case where the various indicators have greatly different scatter.  For example, if a GRB only has a distance modulus from the $V-L$ and $E_{peak}-E_{\gamma}$ relations (i.e., $\mu_2$ and $\mu_4$), then the disparate uncertainties (with $\sigma_{\mu_2} \gg \sigma_{\mu_4}$) will result in the variability contributing little to the final answer.

	A potential problem with equation 29 arises if the luminosity relations are correlated.  In an extreme case, where two relations are perfectly correlated, the use of equation 29 would incorrectly double the weight of what is really just one relation, which is equivalent to reducing the error bars by a factor of 1.4.  This would be non-optimal but not disastrous.  A less extreme example of this would be if the $E_{peak}-E_{\gamma.iso}$ relation of Amati et al. (2002) was used at the same time as the $E_{peak}-E_{\gamma}$ relation, as much the same physics and input are used in both.  For application to this paper, I wonder whether the $E_{peak}-L$ and $E_{peak}-E_{\gamma}$ relations are significantly correlated as they share a common input of $E_{peak}$.  To test this, I have tabulated $\mu_i - \mu_z$ for all bursts and relations, where $\mu_z$ is the distance modulus derived from the observed redshift and some fiducial cosmology (here, the concordance cosmology).  I find that the $E_{peak}-L$ and $E_{peak}-E_{\gamma}$ relations have a near-zero correlation coefficient, which demonstrates that the two relations are not correlated and hence independent.  However, there is one significant correlation, with the correlation coefficient equalling 0.53 between the $V-L$ and $\tau_{RT}-L$ relations.  This is not surprising, as the V of a light curve is dependent on the rise time, and indeed the minimum rise time was originally proposed as a measure of 'variability' that has a direct physical interpretation.  From the extent of the $\mu_2-\mu_z$ versus $\mu_5-\mu_z$ scatter plot, the correlation is roughly half of the intrinsic scatter.  For the 71\% of the bursts with both $V$ and $\tau_{RT}$, the statistical weight of the $V-L$ relation should be roughly cut in half, which is equivalent to the combined $\sigma_{\mu_{2,5}}$ being 20\% too small.  These two relations provide relatively little weight for well-observed bursts (with all five luminosity indicators known), so that the $\sigma_{\mu}$ will only be 4\% too small in these cases.
	
	The statistical weight of each measured distance modulus is $\sigma_{\mu_i}$, so the total statistical weight for each relation contributing to the HD will be a summation of $\sigma_{\mu_i}$ over all bursts.  This will give us an idea of the relative contribution of each of the relations.  I find the summed weights to be 30.3, 27.0, 54.2, 64.9, and 38.4 for the $\tau_{lag}-L$, $V-L$, $E_{peak}-L$, $E_{peak}-E_{\gamma}$, and $\tau_{RT}-L$ relations respectively.  This corresponds to percentages of 14\%, 13\%, 25\%, 30\%, and 18\% respectively.  The $E_{peak}-E_{\gamma}$ relation is the most accurate of the five, but it does not dominate since only 27 bursts are useable.  The $V-L$, $E_{peak}-L$, and $\tau_{RT}-L$ relations all have information for almost all bursts, so their percentages is an indication of the size of the derived uncertainties.  In all, we see that all five relations have comparable total weight, with none dominating and none being negligible.

	Each of these measures carries information, so it would be wrong to not use them with appropriate error bars.  Optimally, we want to use all information, so that the use of any one indicator is neglecting the majority of the constraints.  For example, $\sigma_{\mu}$ will improve from typically 0.7 mag if only the $E_{peak}-E_{\gamma}$ relation is used to around 0.4 mag if all five relations are used.
	
	To illustrate this calculation and some intermediary values, I present the various values for $\mu$ in Table 6.  The first column is the GRB identifier and the second gives the redshift.  Columns 3--7 are the $\mu_i \pm \sigma_{\mu_i}$ values all for the concordance cosmology case with $\Omega_M = 0.27$ and $w=-1$.  In column 8, I give the derived distance modulus from equations 25 and 26.  For comparison, column 9 gives the derived distance modulus for the case of Riess' cosmology ($w_0 = -1.31$ and $w\arcmin = 1.48$).  In comparing the last two columns, there are shifts in the derived $\mu$ values depending on the assumed cosmology, in this case with a typical shift of 0.25 mag with only a small and loose dependence on redshift.

\section{Hubble Diagram}

	The GRB Hubble diagram is a plot of the distance moduli versus the redshifts.  Table 6 has already listed $\mu$ and $z$ values for two cosmologies.  As such, Figures 7 and 8 are GRB HDs as taken straight from Table 6.
	
	Figure 7 shows the GRB HD for 69 GRBs out to redshift $>$6 for the concordance cosmology (indicated by the curved line).  Several aspects of this Figure are striking:  First, the GRBs define a well-behaved curve.  Second, the regime covered by the supernovae ($z<1$ except for 10 events with $1<z\lesssim 1.7$) is only a fraction of the left hand side of this plot.  Third, about half the points are at $z>2$ with good coverage spanning a redshift range far higher than that available even with satellite observations of supernovae.  Fourth, the observed data points are in good agreement with the model curve (with a reduced chi-square of 1.05).  Fifth, the implication is that the GRB HD is consistent with $w=-1$ and an unchanging Dark Energy.
	
	Figure 8 shows the HD for the same 69 GRBs except that the luminosity relations were calibrated with a particular cosmology that has the Dark Energy changing with time.  The goal here is to illustrate the relative lack of shifting of the $\mu$ values as the cosmology parameters change over a typical range.  Between Figures 7 and 8, the points shift by an average of 0.23 mag (but this constant shift does not affect the {\it shape} of the experimental HD curve) while the RMS scatter in the shift is only 0.10 mag largely independent of redshift.  These small relative shifts are to be compared to the change in the model $\mu$ by 0.65 mag at $z=4$.  The conclusion is that the position of GRBs in the HD is independent of the input cosmology (over a reasonable range of parameters) to first order.  The particular cosmology in Figure 8 is that of the best fit by Riess et al. (2004) as based on their 'gold sample' of supernovae.  Riess' cosmology is seen to lie significantly below the data for $z>3$ and above the data for middle redshifts of $1<z<3$.  This situation can be quantified with a chi-square parameter comparing the observations and the model.  Riess' cosmology gives a chi-square of 80.3 while the concordance cosmology gives  a chi-square of 72.3.  This difference of chi-square implies a nearly three sigma rejection of Riess' cosmology when compared to the concordance cosmology.
	
	An important point is that the luminosity relations have to be calibrated for every set of cosmological parameters under consideration.  Schematically, this is different from the supernova HD where the calibration of the decline rate versus luminosity relation can be done at low redshift independent of cosmology and then applied to high redshift events.  All GRB distances are dependent on the cosmology because there is no low-redshift set to perform the cosmology-independent calibration.  This forces us to make a simultaneous fit of the cosmology and the luminosity relations (Schaefer 2003a).  This simultaneous fitting allows us to completely avoid any circularity.
	
	The procedure of calibrating the luminosity relation for each cosmology ensures that the constructed HD has the deviations between observation and model being near zero.  The vertical positioning of the GRB points in the HD are thus 'normalized' out simply by increasing or decreasing the adopted luminosities.  But this is fine, since it is the {\it shape} of the HD that provides the cosmology information.  As the cosmological parameters vary, the deduced $d_L$ for each burst will vary, the positions in the calibration plots (Figures 1-5) will shift, the fitted calibration equations (equations 12, 14, 16, 18, and 20) will change, and the derived distance modulus in the HD will shift.  In the change from the concordance to the 'Riess' cosmology, the model curve has moved downward with a shift that increases to higher redshift.  As such, the $d_L$ value for each burst is lower for the 'Riess' cosmology and so the intercept in all the luminosity relations will decrease.  An overall lowering of the intercepts will not change the {\it shape} of the HD.  So what matters are the shifts relative to this overall offset.  Fortunately, there is a strong mixture of burst luminosities and redshifts, as this avoids a degeneracy where the luminosity relations and the cosmology cannot  become untangled.  The fact that the highest redshift GRBs can only be high luminosity implies that there will be some correlations.  The typical degree of this correlation has already been seen in the comparison of the $\mu$ values between the concordance and 'Riess' cosmologies.  Here, the average scatter in the shifts (0.10 mag) is a small fraction of the shift in the model (0.65 mag at $z=4$), while the systematic shifts with $z$ are also relatively small (0.1 mag from redshifts 1 to 4).  Another way to show that there is a good mixing between redshift and luminosity is that the median redshifts in the ranges $51\leq \log L < 52$, $52\leq \log L < 53$, and $53\leq \log L$ are 1.57, 2.42, and 1.48 respectively.  It is this mixing of redshifts and luminosities that allows for the luminosity relations to distinguished from the cosmology during a simultaneous fit.
	
	What is the accuracy of a single GRB in the HD?  One measure of this is the median $\sigma_{\mu}$, which is 0.60 mag.  (The sizes of $\sigma_{\mu}$ values are apparently reliable since the reduced chi-square for the HD is near unity.)  Another measure is that the deviations have an RMS scatter of 0.65 mag.
	
	A measure of the accuracy in the GRB HD is to plot my derived redshifts (based on the five luminosity indicators) versus the known redshifts (based on optical spectroscopy).  This is shown in Figure 9.  The scatter of the derived redshifts about the diagonal line (where the derived redshifts equal the known redshifts) has a reduced chi-square near unity, and this means that the quoted error bars are realistic.  If we look at the differences between the derived and known redshifts, the RMS error is 26\%.  The maximum error is 69\%, with the second worst error being 56\%.  One the good side, 15\% of the Swift bursts have 14\%-18\% fractional errors in redshift.
	
	The number of degrees of freedom in a chi-square comparison of the GRB data and theoretical models depends on the question being asked.  If we want to do an evaluation of the concordance cosmology, then there are no free parameters for fitting.  (The parameters for fitting the luminosity functions do not count as we are not optimizing on the goodness of fit in the HD.  If we instead let the calibration parameters vary so as to optimize the fit in the HD, then we get correspondingly lower chi-squares in the HD yet poor calibration fits.)  That is, we just accept $\Omega_{M}=0.27$, $\Omega_{\Lambda}=0.73$, and $w=-1$ and we have no changeable parameter to optimize the goodness of fit in the HD.  In this case, the number of degrees of freedom is just the number of bursts in the GRB HD (69).  If we are seeking to constrain cosmological parameters, then the number of degrees of freedom will be the number of bursts (69) minus the number of adjustable parameters.  So if we take a flat universe with $\Omega_{M}=0.27$ as a prior while asking for the best values of $w_0$ and $w_a$, then we have 69-2=67 degrees of freedom.  If we test a more general model where $\Omega_M$, $\Omega_{\Lambda}$, $w_0$, and $w_a$ are allowed to vary so as to optimize the fit in the HD, then we have 65 degrees of freedom in the chi-square.
	
	Two further tasks need to be accomplished before the GRB HDs, such as presented in Figures 7 and 8, can be used to constrain cosmological parameters.  First, we have to evaluate biases arising from gravitational lensing and Malmquist effects.  Second, we have to run chi-square fits over ranges of parameters and marginalize 'nuisance' parameters.
	
\section{Gravitational Lensing and Malmquist Biases}

	The light from high redshift GRBs travels a long way in our Universe, and there is significant probability that it will pass close enough to a galaxy to be gravitationally lensed.  Infrequently, this lensing will magnify the apparent brightness of the burst as seen here at Earth, perhaps with high amplification.  More commonly, lensing will demagnify the image and make it appear somewhat fainter.  The average magnification must be exactly unity.  This magnification and demagnification can cause scatter in the observed HD.  To see this, imagine some particular GRB at some particular redshift with some particular luminosity indicators.  In the case with no lensing, the luminosity indicators would return the approximately correct luminosity and we would use the observed peak flux to return nearly the correct luminosity distance and $\mu$ for plotting on the HD.  In the case with lensing, neither the observed redshift nor the luminosity indicators will be any different, but the observed peak flux will have changed by some factor.  With a shifted peak flux, the deduced luminosity distance and $\mu$ will be shifted and the GRB will be plotted on the HD with a somewhat shifted position.  Most bursts will appear slightly fainter than then they should if unlensed, but some will appear brighter and rarely they can appear much brighter.  There is no observational means of measuring the magnification for any GRB.  The result will be an increased and irreducible scatter in the HD, and indeed this expected scatter is likely to account for some of the systematic scatter observed in the calibration plots (Figures 1-5).
	
	Several groups have calculated the probability distributions for magnifications as a function of redshift.  Holz \& Linder (2005) calculate that the effective RMS of the scatter from lensing is roughly $0.093z$ mag up to $z=2$.  They find that with 70 sources at $z=1.5$ the lensing effects will average out to errors of less than one percent.  Premadi \& Martel (2005) present probability distributions for magnifications up to a redshift of 5, and find that the dispersion turns over for $z>1$ so that the Holz \& Linder result should not be used for $z\gtrsim1$.  Oguri \& Takahashi (2006) make a first attempt to calculate the effects of lensing on a GRB HD, and they conclude that the lensing effects are negligible.  All three results point towards lensing not being a significant bias for the GRB HD.
	
	The effects on the HD are more complicated than just the addition of some scatter in the deduced $\mu$.  The reason is that there is effectively a threshold in burst apparent brightness for its being detected, its redshift being measured, and luminosity indicators being measured.  Although it is a 'fuzzy' limit, there will be a corresponding distance limit beyond which GRBs will not be included in the sample in Table 2.  With gravitational lensing, bursts just outside this threshold might be magnified in brightness and included in this sample, whereas bursts just inside this threshold might be demagnified in brightness and excluded from this sample.  This threshold effect will result in the highest redshift events appearing systematically more luminous than they would be in the absence of lensing.  The quantitative evaluation of these effects depends on the shape of the luminosity function (Oguri \& Takahashi 2006).
	
	There is a similar set of effects (referred to generically as Malmquist bias) that result from observational uncertainties and geometry effects near the threshold of detection.  With observational errors and intrinsic scatter in the luminosity relations, bursts just inside the distance limit will be excluded if the random fluctuations push the apparent brightness below the detection threshold, while bursts just outside the distance limit will be included if the random fluctuations push the apparent brightness above the detection threshold.  There is more volume just outside the distance limit than there is just inside, so this will result in more over-bright GRBs being included than under-bright GRBs being excluded.  A similar effect  occurs for the luminosity because there are always many more lower-luminosity events that can have random fluctuations excluded in the sample than there are higher-luminosity events that can have random fluctuations included in the sample.
	
	Gonzalez \& Faber (1997) provide a general expression that can be used to calculate both the gravitational lensing biases and the Malmquist biases.  In general for any particular GRB, the expression for the bias is
\begin{equation}
(1+B)D = \langle r|D \rangle = \frac{\int_0^{\infty} rP(r|D)dr}{\int_0^{\infty} P(r|D)dr} =  \frac{\int_0^{\infty} rP(r,D)dr}{\int_0^{\infty} P(r,D)dr} .
\end{equation} 
$D$ is the 'raw' or derived distance to a source that is really at distance $r$.  $B$ is the bias in the derived distance $D$.  The bias in the distance modulus will be $\mu_{bias}=5\log (1+B)$, and this is what is sought in this section.  The one-sigma error bar for $\mu_{bias}$ is calculated with the usual higher moments of $r|D$.  The expectation value of $r$ given $D$ is written as $\langle r|D \rangle$.  $P(r|D)$ is the conditional probability of $r$ given $D$.  $P(r,D)$ is the joint probability which can be separated as
\begin{equation}
P(r,D)=P(r)P(D|r) = r^2n(r)P(D|r) .
\end{equation} 
Here, P(r) is the probability of GRBs (observed or unobserved) being at a true distance of $r$, with this being $r^2n(r)$ where $n(r)$ is the rate density of GRBs.  The $r^2$ term accounts for the increased volume at greater distances.  $P(D|r)$ is the probability of measuring a distance $D$ given that the burster is really at distance $r$.  This probability contains all the information about the measurement uncertainties, the intrinsic scatter in the luminosity relations, the GRB luminosity function, and the gravitational lensing probabilities.

	The probability of measuring a distance $D$ when the real distance is $r$, i.e., $P(D|r)$, depends on many factors.  In general it can be expressed as 
\begin{equation}
P(D|r)=[P_M \ast (P_L G)]P_{sample},
\end{equation} 	
for GRBs.  Here the $\ast$ symbol indicates a convolution.  $G$ is the Gaussian distribution of the measurement errors.  $P_L$ is from the burst luminosity function.  The product $P_L G$ represents that there are more low-luminosity than high-luminosity bursts, so that deviations in that direction are more likely.  $P_M$ is the probability distribution of lensing with magnification $M$.  The convolution in equation 32 produces the probability distribution of the burst brightness.  The multiplication by $P_{sample}$ is to account for the practicalities of detection of a GRB and its subsequent redshift measure as required for inclusion in this sample.

	The intrinsic scatter in the luminosity relations and the results of measurement errors can both be modeled together as  a Gaussian distribution in log-space.  This log-normal distribution is 
\begin{equation}
G(r,D)=e^{-\frac{1}{2} [(\ln r - \ln D)/\sigma_{\ln D}]^2}.
\end{equation} 
The distance $D$ is derived from the luminosity indicators (to get $L$), the observed peak flux ($P_{bolo}$), and equation 23.  The width of this distribution is given by
\begin{equation}
\sigma_{\ln D}=0.5[(\sigma_L/L)^2+(\sigma_{P_{bolo}}/P_{bolo})^2]^{0.5}.
\end{equation} 
The luminosity $L$ here is from the weighted average of the luminosities derived from the luminosity indicators (equations 12, 14, 16, 18, and 20).  The uncertainty in that luminosity ($\sigma_L$) is from equations 13, 15, 17, 19, and 21 except that the systematic errors are set equal to zero.  The reason is that the systematic errors likely arise from the lensing and Malmquist effects so it would count them twice to include them in the $\sigma_L$ values here.

	Firmani et al. (2004) derive the luminosity function for GRBs based on the peak flux distribution of BATSE bursts plus  33 GRBs with measured redshifts.  The simplest acceptable models have the probability of a burst having a luminosity $L$ as 
\begin{equation}
P_L \propto L^{-1.57}.
\end{equation} 
The constant of proportionality does not matter as it divides out in equation 30.  This can be translated (with equation 23) into a dependence on distance as
\begin{equation}
P_L(D) \propto r^{-3.14},
\end{equation} 
for any individual burst.

	The lensing magnification $M$ is the factor by which the unlensed brightness is multiplied to get the observed brightness of the burst.  The average value of $M$ must be unity.  The majority of bursts will have $M<1$ while there will be a small tail to high $M$.  To take a specific example, for $z=3$, the smallest value of $M$ will be 0.81, the most likely value is 0.89, the probability distribution has fallen to 10\% of its peak at $M=0.82$ and $M=1.16$, and there is a 1\% chance of having $M>1.9$ (Premadi \& Martel 2005).  The probability distribution of any particular magnification, $P_M$, will depend on redshift.  I have adopted the calculated $P_M$ curves out to $z=8$ from Premadi \& Martel (2005).

	The probability of including a GRB in my sample of 69 bursts ($P_{sample}$) depends on many factors which are poorly known and time variable.  The detection of a particular burst requires some satellite experiment to be looking in the right direction at the right time to a threshold fainter than the burst (given its light curve shape).  To a first approximation, the burst will be detected if its peak flux is brighter than the experimental threshold, which varies from satellite to satellite.  Not only must the burst be detected, but it must also have a redshift determined from optical spectroscopy, and this depends on many variables relating to observing conditions and followup programs.  To a first approximation, this probability depends on the brightness of the optical afterglow, which crudely scales as the peak flux of the burst.  (Indeed, the Swift bursts have a probability of getting a redshift being only weakly dependent on the peak flux; with 28\%, 38\%, and 39\% of the faintest-third, middle-third, and brightest-third, respectively, of the bursts discovered by Swift having redshifts.)  So a reasonable approximation is to have $P_{sample}$ be only a function of $P_{bolo}$.  To quantify this, I have constructed a $\log N_{>P}-\log P_{bolo}$ distribution for the GRBs in my sample, where $N_{>P}$ is the number of bursts brighter than any given $P_{bolo}$.  I have broken the sample into bursts according to the primary discovery satellite, as this will lump together GRBs discovered with a constant detection threshold and with similar dynamics in the optical followup community.  With a knowledge of these thresholds and how followup efforts have changed over time, I have further grouped the BATSE and Konus discoveries together and well as the SAX, Integral, and HETE-2 events together.  The resultant $\log N_{>P}-\log P_{bolo}$ curves all have a steep region that is roughly -3/2 in slope which represents a uniform $P_{sample}$ above some threshold $P_{threshold}$.  The curves all break at some peak flux and roll over in a manner well-described as a Gaussian.  A Gaussian curve will have a long tail that must be cutoff somewhere, so I have adopted $z=7$ as a cutoff due to the problems at getting spectra far enough into the infrared for higher redshifts.  This probability function will be normalized to be unity for bright bursts, with this normalization choice dividing out in equation 30.  Then, 
\begin{eqnarray}
P_{sample} = \left\{ \begin{array}{lll}
	 1 & \mbox{if $P>P_{threshold}$} \\
	 e^{-\frac{1}{2} [\log (P_{bolo}/P_{threshold})/W]^2} & \mbox{otherwise}\\
	 0 & \mbox{if $z>7$}.
	 \end{array}
\right. \
\end{eqnarray} 
Here, $W$ is a constant representing the sharpness of the cutoff in the discovery probabilities.  For all satellites, I find that $W=0.5$ fits well with the observed distributions.  The $P_{threshold}$ values I find are $10^{-5}$ erg cm$^{-2}$ s$^{-1}$ for BATSE and Konus bursts, $10^{-6}$ erg cm$^{-2}$ s$^{-1}$ for SAX, Integral, and HETE-2 bursts, and $3 \times 10^{-7}$ erg cm$^{-2}$ s$^{-1}$ for Swift bursts.  For use in equation 32, the peak flux has to be translated into a distance $D$ for the given burst.  As I am looking at brightness distributions for bursts in my sample, my formulation simultaneously approximates the probabilities for both the burst detection as well as the redshift measurement.
	
	The number density ($n(r)$) of GRBs changes changes with distance, at least roughly proportional to the average high-mass star formation rate.  Firmani et al. (2004) derive typical number densities as a function of redshift;
\begin{equation}
n(z) \propto e^{2.3z+c(z-2)}/(e^{2.3z}+80),
\end{equation} 
where $c=0$ for $z<2$ and $c=0.25$ for $z>2$.

	After all this, we can evaluate the bias with
\begin{equation}
(1+B)D =\frac{\int_0^{\infty} r^3n(r)[P_M \ast P_LG]P_{sample}dr}{\int_0^{\infty} r^2n(r)[P_M \ast P_LG]P_{sample}dr} .
\end{equation} 
This equation is to be applied to individual bursts.  The required input is the satellite (to set $P_{threshold}$), $z$ (to select  the $P_M$ curve), $L$ and $P_{bolo}$ (to determine $D$), as well as $\sigma_L$ and $\sigma_{P_{bolo}}$ (to determine $\sigma_{\ln D}$).  

	Let me discuss the various competing effects to try to give some insight into the size and direction of the bias.  There are two effects that will tend to make any given burst have a smaller measured distance than the GRB really is (i.e., $D<r$).  The first is that there is greater volume at further distances so more GRBs will be included from distances greater than $D$ than excluded from distances less than $D$.  This is represented in equation 39 by the $r^2$ factor.  The second is that GRBs are more common as we go to higher redshifts, so there are more GRB to be included with $D>r$ than there are to be excluded with $D<r$.  This is represented in equation 39 by the $n(r)$ factor.  There are also two effects that work in the opposite direction.  The first is that if a burst appears farther than it really is ($D>r$), then it is less likely to be detected and a redshift measured.  This is represented in equation 39 by the $P_{sample}$ factor.  The second effect is that there are more GRBs with lower luminosity so that there will be more GRBs with $D<r$ to get included in this sample than there are higher luminosity events to get excluded.  This is represented in equation 39 by the $P_L$ factor.  Thus we have two effects that are in competition with two other effects determining whether the bias will be positive or negative.  The measurement error distribution ($G$ in equation 39) has an average of zero, so this term does not introduce any bias.   The lensing magnification distribution ($P_M$ in equation 39) has an mean of unity so that it also does not introduce any bias on average, even though it does provide scatter in the HD.

	The biases for my 69 burst sample have been calculated and presented in Table 7.  Columns 1-3 are the usual GRB name, the detecting satellite, and the redshift.  Column 4 is the peak luminosity (with units of erg s$^{-1}$) as calculated from the luminosity indicators along with its uncertainty with no contribution from systematic errors.  Column 5 is the bolometric peak flux and its uncertainty in units of erg cm$^{-2}$ s$^{-1}$.  Column 6 is the derived combined gravitational lensing and Malmquist biases expressed as a shift in the distance modulus, $\mu_{bias}$ and its uncertainty in units of magnitudes.

	In the absence of gravitational lensing, accurately measured distances (i.e., with small $\sigma_L$ and $\sigma_{P_{bolo}}$) will result in very small biases.  Indeed, the bias will scale as the square of $\sigma_{\ln D}$ (e.g., Gonzalez \& Faber 1997).  For GRBs, $\sigma_{\ln D}$ is not small, so the first expectation is that the biases will be quite large.  But a glance down the last column in Table 7 shows that the biases are quite small.  The average value of $\mu_{bias}$ is 0.03 mag while it has an RMS scatter of 0.14 mag.  Thus the biases are small compared to the the typical error bars in $\mu$ as well as the typical changes in $\mu$ for the model HD at moderate redshifts.  There are two reasons why the GRB biases are small.
	
	The first reason for small biases is that the bursts with redshifts are on average the same ratio above the experimental threshold.  That is, the faint bursts were discovered with satellites with low thresholds while the bright bursts were discovered with experiments with high thresholds.  Thus the median values of $P_{bolo}/P_{threshold}$ is 0.87 for BATSE and Konus, 0.52 for SAX, HETE-2, and Integral, and 1.08 for Swift -- despite the thresholds varying by over a factor of 30.  This results in the values of $P_{bolo}/P_{threshold}$ having no significant trend with redshift; with median values of 1.09 from $0<z<1$, 1.27 from $1 \le z<3$, and 0.64 from $3\le z<7$ -- despite the $P_{bolo}$ changing by a factor of 1200.  That $P_{bolo}/P_{threshold}$ is largely independent of redshift causes the bias corrections to also be largely independent of redshift, and thus that the shape of the HD does not change.
	
	This first reason for the low biases for GRBs is identical to one of the reasons why the Malmquist bias is claimed to be negligible for supernovae in the HD (Perlmutter et al. 1997; 1999; Knop et al. 2003).  In this case, the low redshift sample of supernovae (from Hamuy et al. 1996) had relatively bright peak magnitudes yet the discoveries were made with a relatively high detection threshold, while the high redshift sample supernovae had relatively faint peak magnitudes yet the discoveries were made with a relatively low detection threshold.  The logic was that both samples had similar brightnesses above thresholds at the time of discovery so that any Malmquist bias would shift the two samples by similar amounts resulting in no change of shape to the HD.
	
	The second reason for the small biases for GRBs is that the product $r^2n(r)P_LP_{sample}$ is largely flat for most of the GRBs in my sample.  That is, fortuitously, most bursts have that product roughly constant across the likely range of distances.  In this case, the integrals in equation 39 will simply return $D$, so that both $B$ and $\mu_{bias}$ will be near zero.  Let me illustrate this with a specific case of how the product changes from $z=1$ to $z=2$.  The $r^2$ term increases by a factor of 5.5, the $n(r)$ term increases by a factor of 16, the $P_L$ term decreases by a factor 0.24, and the $P_{sample}$ term decreases by a factor of 0.21 for GRBs farther than the threshold.  In all, the product $r^2n(r)P_LP_{sample}$ changes by a factor of 0.44 across this range in redshift.  And most bursts have their likely range of redshift being much smaller than this, so that the change in the product across their likely distances will generally be much smaller resulting in a small $\mu_{bias}$.
	
	This second reason is similar to the case for small Malmquist effects for the supernova HD (Perlmutter et al. 1997; 1999; Knop et al. 2003).  These papers explicitly cite the offsetting effects that cancel out as part of their justification that the Malmquist bias can be neglected.
	
	In all, we see that the gravitational lensing and Malmquist biases are much smaller than the quoted error bars and smaller than typical variations in the models.  Even though the systematic biases are low, there is still substantial scatter introduced into the GRB HD by these two effects.  At a redshift of 3, the RMS scatter in $\mu$ due to variations in the lensing magnification is roughly a quarter of a magnitude (Premadi \& Martel 2005), and this is comparable to the systematic scatter in the luminosity relations.  Thus, I attribute the 'extra' scatter to the lensing and Malmquist effects.

\section{Cosmological Constraints}

	To put constraints on parameters in cosmological models, we must first adopt a set of priors as constraints on allowed values for parameters and we must identify which parameters are statistical 'nuisance parameters' (whose value we do not care about) for marginalization.
	
	Different workers will choose different priors.  Throughout this paper, I will take the universe to be flat ($\Omega_M+\Omega_{\Lambda}=1$).  For the cases where I am examining possible variations in the Dark Energy, I will adopt priors of either $0<\Omega_M<1$ or $0.23<\Omega_M<0.31$.
	
	The identity of nuisance parameters depends on the science question being asked.  If we are interested in constraining the cosmological parameters, then we really do not care about the value for the slope of the $V-L$ relation (or any of the other nine fitted constants in the luminosity relations).  The process for marginalizing these nuisance parameters is to integrate over them.  For example, if we assumed a flat universe and $w=-1$ and are only interested in $\Omega_M$, then the probability for a particular value of $\Omega_M$ is given by 
\begin{equation}
P(\Omega_M) = \int \int \ldots \int P(\Omega_M,~a_1,~b_1, \ldots , b_5) da_1db_1 \ldots db_5,
\end{equation} 
where $a_1$ and $b_1$ are the intercept and slope of the first luminosity relation, and so on.  These probabilities are related to the observed chi-squares as
\begin{equation}
P \propto e^{-\chi^2/2}.
\end{equation} 
If we are interested only in the Dark Energy parameters, then we might consider $\Omega_M$ as a nuisance parameter that also needs to be marginalized.
	
	What are the best fit parameters for the cosmological models?  Here, I will examine three separate questions:
	
	If Dark Energy is constant with $w=-1$ and the universe is flat, then what is the best value for $\Omega_M$ from GRBs alone?  This is somewhat of a test case, as supernovae and other evidence already tell us that $\Omega_M=0.27 \pm 0.04$ in this case.  Also, other GRB studies have results that can be used for this same question.  Thus, Firmani et al. (2006) find $\Omega_M = 0.29_{-0.06}^{+0.08}$ while Xu, Dai, \& Liang (2005) find $\Omega_M = 0.30_{-0.06}^{+0.09}$.  Figure 10 presents $P(\Omega_M)$ with an arbitrary normalization.  I find that $\Omega_M = 0.39_{-0.08}^{+0.12}$.
	
	What constraints can we place on the Dark Energy parameters ($w_0$ and $w_a$) if we are not interested in the value of $\Omega_M$ in a flat universe?  For this question, we have to marginalize over the luminosity relation parameters as well as $\Omega_M$ over the range 0--1.  Figure 11 shows the resulting one-sigma and two-sigma confidence regions.  The best fit set of parameters is $w_0=0.1$ and $w_a = -0.9$.  If we are only interested in one of these parameters, then the one-sigma uncertainty range, then this will be spanned by the one-sigma region in Figure 11.  Thus, $w_0=0.1_{-0.5}^{+0.4}$ and $w_a = -0.9_{-1.2}^{+1.0}$ at the one-sigma level and $w_0=0.1_{-1.3}^{+1.5}$ and $w_a = -0.9_{-5.1}^{+2.1}$ at the two-sigma level.  But these simple one dimensional error bars do not represent the correlation in Figure 11.  So perhaps a better representation of the one-sigma confidence region would be
\begin{equation}
w_a + 2.2 w_0 = -0.68^{+0.25}_{-0.48},~~
w_a - 0.45 w_0 = -0.95^{+1.19}_{-1.40}.
\end{equation}
In the thin dimension, the width of the one-sigma region is 0.73.
	
	In this case, the Cosmological Constant ($w=-1$) has a marginalized probability that is only 12\% of the peak probability, and this puts that position just outside the two-sigma confidence region.  However, the best fit area on this diagram corresponds to a relatively shallow slope for $z<0.5$, and this is inconsistent with the supernovae data.  If the supernova data is used (either as a prior or as part of a joint fit) to exclude much of the region substantially  below and to the right of the Cosmological Constant point, then the Cosmological Constant point will likely go inside the one-sigma confidence contour.  Another reason to not reject the Cosmological Constant from this analysis is that the chi-square for the best fit luminosity relations (i.e. without any marginalization) is 70.7 (for 66 degrees of freedom) for the best point in Figure 12 and is 72.3 (for 69 degrees of freedom) for the Cosmological Constant point.  Naturally, the use of three additional fit parameters has improved the chi-square, but an F-Teast shows that the improvement is much too small to justify the use of  the additional parameters.  Thus, the improvement in chi-square is too small to reject the Cosmological Constant.  Another way to see that the Cosmological Constant should not be rejected is simply to recall that the reduced chi-square for its HD fit is 1.05, and this indicates an acceptable model.
	
	If we accept the prior work yielding $\Omega_M=0.27 \pm 0.04$ in a flat universe, then what are the limits on the Dark Energy parameters?  For this question, we have to marginalize over the luminosity relation parameters as well as $\Omega_M$ over the range 0.24 -- 0.31.  Figure 12 shows the resulting one-sigma and two-sigma confidence regions.   The case with a closely restricted $\Omega_M$ has a best fit location of $w_0=0.2 \pm 0.6$ and $w_a = -1.4 \pm 1.2$ (at the one-sigma confidence level), while its unmarginalized chi-square is 70.7.   The one-sigma confidence region can be represented by
\begin{equation}
w_a + 2.1 w_0 = -1.0 \pm 0.3,~~
w_a - 0.48 w_0 = -1.5 \pm 1.5.
\end{equation}
In the thin dimension, the width of the one-sigma region is 0.6.  The comparison between Figures 11 and 12 shows that they are largely the same.  Nevertheless, the two-sigma region for Figure 11 is somewhat larger than in Figure 12, as appropriate for its looser constraint on $\Omega_M$.
	
	In the case with the more restrictive prior on $\Omega_M$, the Cosmological Constant ($w=-1$) has a marginalized probability that is only 11\% of the peak probability, and this puts that position just outside the two-sigma confidence region.  Again, the results of this fit do not allow for the rejection of the Cosmological Constant.
	
	In both Figures 11 and 12, we find the Cosmological Constant point being just outside the two-sigma confidence region.  But we cannot reject the Cosmological Constant on this basis, as discussed above.  The real bottom line is that the simple concordance case ($w=-1$ and $\Omega_M=0.27$ in a flat universe) provides a nice fit to the data (see Figure 7) as evidenced by a fine reduced chi-square of 1.05.  By allowing three additional parameters to vary, we can always get a slightly better reduced chi-square, in this case with an improvement from 72.3 to 70.7.  In all, I find that the GRB HD is fully consistent with a constant Cosmological Constant.

\section{How Robust is the GRB HD?}

	How robust are the results coming from the fits to the GRB HD?  That is, for reasonable differences in various choices in the analysis, will the chi-square contours (as in Figures 11 and 12) change significantly?  I can perform a variety of 'sanity checks' and I can make a variety of procedural changes to see the sensitivity of the results to the changes.  In this section, I will try seeking for systematic changes in the HD and fits as various selections are varied.
	
	The results are tabulated in Table 8, with the various cases specified and the number of GRBs used in each case.  To quantify the confidence contours, I can determine the minimum chi-square, the position of the long axis and the width perpendicular to that axis (expressed as a one-sigma error bar on that position), the acceptable position along that axis (expressed as a range of $w_0$), the chi-square for the concordance model, and the chi-square for the 'Riess' model.  The first line in Table 8 is for the 'standard' data set from Tables 2-4, corresponding to the case displayed in Figure 12.  Subsequent lines are for different cases and should be compared back to the first line to see whether the differences are significant.
	
	The HD has only two GRBs with $z>5$, so we can wonder how sensitive the results are to these two points?  That is, are the conclusions dominated by just two bursts, or did biases in these two events significantly shift the cosmological constraints?  To address these questions, I have repeated the entire fitting of the GRB HD (as in Figure 12) while deleting GRB 050904 and GRB 060116 from the data set.  We see the that the chi-squares are somewhat smaller due to eliminating two bursts (as expected).  Also, the chi-square differences between the various models is lower, as expected when the two points with the highest redshifts are arbitrarily ignored.  The position of the confidence contours are essentially unchanged.  In all, the constraints on cosmology and the conclusions are not affected by a deletion of the two highest-redshift bursts.
	
	How sensitive are my cosmological fits to the rejection of the outliers as itemized in Section 3.   Recall, that I rejected three bursts and two rise times as being outliers.  One way to test this is to repeat the fitting to the GRB HD except to include all the rejected input.  The properties of the new chi-square contours are tabulated in Table 8.  We see that the minimum chi-square is very large (209.9 versus 70.7 with the outliers rejected), but this is exactly as expected.  Most of this large increase comes from just the one burst GRB 980425.  The central axis of the long-thin confidence contours is the same, although the high position on the HD and low redshift of GRB 980425 pushes the minimum chi-square to $w_0 > 2$ along this axis.  This is not surprising, as any one very distant outlier can force a substantial tilt in the best fit and hence push the contours around.  If we include the data for four of the outliers not including GRB 980425 (i.e., GRBs 020819 and 050315 as well as the rise times for GRBs 990123 and 030328), then we get the contour as shown in Table 8.  The chi-squares are still high when compared to the standard case (as expected).  But largely we see that the position of the contours and the relative chi-squares are not affected by the inclusion of the four outliers.
	
	In Section 2, I rejected five GRBs which have reported "redshifts of too low a confidence to be used", so we can ask whether these rejections are reasonable?  For GRB 980326, the reported redshift of $z\sim1$ (Bloom et al. 1999) was solely based on interpreting a marginally-detected bump in the light curve as being a SN1998bw-like supernova.  However, we now know that the GRB/supernova events span a wide range of peak absolute magnitudes, so the derived redshift must have a very large uncertainty, even assuming that the bump arises from a supernova.  And the $z\sim1$ claim is dubious since the host galaxy would then have to be at least seven magnitudes fainter than L* and this is very unlikely (Fruchter et al. 2001a).  So the reported redshift is much too low in confidence for GRB 980326 to be used in a Hubble diagram.  GRB 011121 has a reported redshift of $z=0.362$ on the basis of a nearby galaxy with no connection to the optical transient position (Garnavich et al. 2003).  Rather, with {\it HST} imaging, the host appears to be a separate 'blue knot' of unknown redshift (Bloom \& Price 2002).  GRB 020305 has a galaxy on top of the optical transient position, for which a low signal-to-noise spectrum yields proposed redshifts of $z\sim<2.8$, $z\sim0.2$, $z>0.5$, and $z=2.5$ (Gorosabel et al. 2005).  Clearly, any of these four possibilities has too low a confidence for GRB 020305 to be used in any Hubble diagram.  GRB 050803 has only an x-ray error circle, and this has "several faint sources in the uncertainty circle", for which the brightest happens to have a redshift of 0.422 (Bloom et al. 2005).  An error circle with 12$\arcsec$ diameter will always have some brightest source and any connection to the GRB can only be conjectural.  GRB 060123 also only has an x-ray error circle and naturally there is some brightest galaxy in that circle (Butler \& Bloom 2006).  A long spectrum from the Gemini 8-m telescope shows only one certain emission line, for which redshifts of $z=1.099$, $z=0.193$, and $z=0.562$ have been proposed (Berger et al. 2006), so that all we have is a galaxy of unknown redshift and unknown connection to the GRB.  In all, the five rejected GRBs have suggested redshifts that are of much too low a confidence for inclusion in any HD work.
	
	A related question is whether any of my 69 GRBs have a redshift of sufficiently low confidence that some would choose to reject those GRBs?  This is partly a difficult question because the available information is very scattered in nature and hard to lead to a quantitative confidence level.  Thus, we cannot simply perform fits as a function of some redshift-confidence threshold.  Instead, I can perform the cosmology fits where I eliminate any of my 69 GRBs for which the redshift has been seriously questioned.  I find that only three of my GRBs have had serious questions raised about their quoted redshifts; GRBs 050802, 060108, and 060116.  GRB 050802 has a spectrum of its optical afterglow with several absorption lines indicating a redshift of 1.71 (Fynbo et al. 2005c), however, McGowan et al. (2005) place a limit of $z<1.2$ based on a UVOT detection in the far ultraviolet.  But the optical redshift is of very high confidence with a good spectrum and many lines (J. Fynbo 2005 private communication), while the redshift of $z=1.71$ is actually consistent with the UVOT detection (K. McGowan 2005, private communication).  As such, I have confidently included this burst into my HD.  GRB 060108 has its redshift based on a fit to the spectral energy distribution of the optical afterglow, with the best value being $z=2.03$ and an upper limit being quoted as $z<2.7$ (Malendri et al. 2006).  As such, perhaps the uncertainty in the redshift is too large for inclusion in the GRB HD?  GRB 060116 has a well-observed optical/infrared afterglow whose spectral energy distribution yields $z=6.60\pm 0.15$ (Grazian et al. 2006).  A more detailed analysis of the same data (Piranomonte et al. 2006) returns the same redshift, but another possibility of $3.8<z<4.5$ is presented with a lower likelihood.  The two possibilities are distinguished by either low extinction (for $z=6.60$) or high extinction (for $z\sim4$).  Piranomonte et al. (2006) report a "very poor" late spectrum with possible light down to 700 nm wavelength, but this is contradicted by the lack of a detection with the {\it HST} ACS/F775W filter (Tanvir et al. 2006), so that the $z=6.60$ value is strongly preferred.  Despite having reasonable confidence for the inclusion of GRBs 050802, 060108, and 060116, we can ask what would be the effect on the cosmology fits if these three bursts are excluded from the analysis.  The resultant chi-square contours are presented in Table 8.  We see that the rejection of the three 'questionable' redshift bursts does not significantly change the cosmology fit results.  In all, the GRB HD appears to be robust to changes in the confidence level for the measured redshifts.
	
	In principle, the weighted average of the distance moduli from all available luminosity relations will combine all information and produce the best value.  But we can worry whether one or more of the relations is adding in more noise than signal.  To test this, I have performed all the fits again while successively eliminating each of the luminosity relations.  The chi-square contour information is in Table 8.  As expected, the minimum chi-square varies around a bit.  However, the primary axis of the confidence region and the relative chi-square values vary little.  An examination of the HDs reveals a well formed set of points following along a smooth curve with no signs of systematic shifts.  As such, I see no evidence that any one relation is adding more noise than signal.  In another trial, I have constructed a HD (see Figure 13) from the standard data set with the elimination of the distance moduli derived from the two relations ($V-L$ and $\tau_{RT}-L$) with the most scatter in the calibration curves.  The plot points fit smoothly along the model curve and all looks fine.  Figure 13 can be directly compared with Figure 7.  As expected with the two  relations carrying 32\% of the total statistical weight, the typical error bars in Figure 13 are substantially larger than those in Figure 7.  In all, there is no evidence that any of the relations is doing more harm than good, so it is best to use all the relations and take advantage of all the available distance information.
	
	If each relation is taken one at a time, then are the resulting chi-square contours consistent with the contours from all five relations?  To test this, I have constructed HDs for each luminosity relation alone.  As expected, the scatter is much larger than displayed in Figure 7.  The contours are presented in Table 8.  The contours are consistent with the standard case, with the expected shifts as will always be seen when looking at a subset of the data.  The largest shift is for considering the $\tau_{lag}-L$ relation.  The $E_{peak}-E_{\gamma}$ relation is surprisingly broad, although it is still consistent with the five-relation contours at the one-sigma level.  In all, I see nothing that points to any one relation being inconsistent with the final result.

	One possible selection effect could arise from the diversity of GRB spectral shapes, especially at higher energies.  A high redshift population of GRBs will have their higher photon energies shifted to the detector's bandpass and then be selected for those with harder spectra (i.e., with less negative $\beta$ values).  In this way, the detected bursts at higher redshifts might have a harder spectrum than the sample of detected bursts at low redshift, even in a case with no evolution of the full GRB population.  I have three reasons for thinking that such effects are minimal:  (A) The high redshift bursts are necessarily of high luminosity and hence (by the $E_{peak}-L$ relation) have a high intrinsic $E_{peak}$.  Schaefer (2003b) has shown that the redshifting comes close to exactly offsetting the intrinsic $E_{peak}$ so that bursts at all redshifts will have the same apparent $E_{peak}$.  (This is actually a function of the observed peak flux, and the insensitivity of of the measured $E_{peak}$ on redshift is the reason that the observed $E_{peak}$ distribution is allowed to be as narrow as it is.)  In this case, the effects of diversity in the GRB spectra on the burst triggering will be identical at both high and low redshifts.  (B) The high energy band contributes only a relatively small fraction of the flux in the trigger range, so variations of $\beta$ will have little effect on burst detection.  To take a specific example, for the usual BATSE trigger range of 50-300 keV, a burst with the median $E_{peak}=300$ keV (Mallozzi et al. 1995) will have zero effects on the trigger caused by varying $\beta$ since the spectral break is above the trigger band.  If $E_{peak}=100$ keV, then a variation of $-2 < \beta < -3$ will only cause the counts in the BATSE trigger range to change by 6\%.  For very low values of $E_{peak}$, say at 30 keV, the effect of varying $\beta$ over such a large range can produce a factor of two variation in the counts in the trigger band.  For the Swift detectors, the trigger band is 15-150 keV, hence making the effects of varying $\beta$ zero for most bursts and small for most of the remainder.  In all, the triggering effect of $\beta$ variations is minimal.  (C) If a selection effect on $\beta$ as a function of redshift exists, then we might expect to see a correlation between $\beta$ and $z$ for the detected bursts.  But such is not seen from bursts with measured $\beta$ values in Table 2.  For example, the weighted averages of $\beta$ for $z<1$ and $z>2$ is $-2.31 \pm 0.05$ and $-2.26 \pm 0.06$ respectively.
	
	A means of pointing to the improvement implemented by the luminosity indicators is to construct a HD with the bursts artificially set to the same luminosity.  This is displayed in Figure 14 for the concordance cosmology.  This figure shows a horrible match with immense scatter.  As such, the comparison between Figures 7 and 14 demonstrates that the luminosity relations are contributing vital distance information.
	
	In summary, the GRB HD and the fits to cosmology appear to be robust.

\section{Comparison with Supernovae}

	For Hubble diagram work, a comparison of GRBs and supernovae is inevitable.
	
	The great advantage of supernovae for the HD is that they are substantially more accurate standard candles than are GRBs.  This will translate into better constraints on cosmological parameters.  But this advantage for supernovae is not as large as some might expect.  The one-sigma scatter of supernovae about the concordance cosmology in the HD is 0.22 mag for the best 31 supernovae out of 42 or 0.36 mag for all 42 (Perlmutter et al. 1999),  0.27 mag for $0.1<z<1$ and 0.29 for $1<z$ even after the selection of the very best supernovae (Riess et al. 2004), and $>$0.25 mag (Astier et al. 2005).  It is clear that the real accuracy of supernovae is $\sim 0.25$ mag only after much selection, so the real accuracy without {\it a posteriori} selection (other than throwing out $>3-\sigma$ outliers) is more like 0.3 mag or 0.35 mag.  This is to be compared to the residuals of my GRB HD where the RMS scatter is 0.65 mag.  So the comparison between supernovae and GRBs is that the scatter in the residuals around the concordance cosmology is 0.3--0.35 mag to 0.65 mag.  That is, supernovae are roughly a factor of two more accurate than GRBs for HD work.  This factor of two is likely to be substantially smaller than many people would expect.  And there is every prospect of significantly improving the GRB luminosity indicators (see Section 10), so this factor is likely to get substantially smaller over the next few years.
	
	A great advantage of GRBs for HD work is that they go out much further in redshift than the many ground-based supernovae ($z<1$) or space-based discoveries from either HST or SNAP ($z<1.7$).  GRBs are already  numerous from $1.7<z \le 6.6$, and it might be possible in the future to get an occasional GRB out to even $z=10$ (Lamb \& Reichart 2000; Fenimore \& Ramirez-Ruiz 2000; Schaefer, Deng, \& Band 2001; Bromm \& Loeb 2002).  The use of high redshift GRBs will provide a long lever-arm for measuring changes in the slope of the HD.  That is, model HDs will have small differences in predicted distance moduli for $z \sim 1$ yet will have substantially larger differences at high redshift (see Figure 15).  For example, the differences between the concordance cosmology and Riess' best fit cosmology is -0.03 mag at $z=0.5$, 0.04 mag at $z=1$, 0.44 mag at $z=3$, and 0.83 mag at $z=5$.  More to the point for a comparison between GRBs and supernovae, the difference for the highest redshift supernova (at $z=1.7$) is only 0.15 mag while the difference for the current highest redshift GRB (at $z=6.6$) is 1.00 mag.  Thus GRBs have a large advantage by using the long lever arm of high redshifts since the models predict large differences at high redshifts.
	
	A common argument is that the HD is uninteresting for redshifts higher than the SNAP range because the universe will be matter dominated and so the Dark Energy will be too small to affect the universal expansion at high redshifts.  But this argument is wrong because it ignores the long lever arm of going to high redshifts in yielding the slope of the HD.  To take a specific example, suppose that we wanted to measure the slope of the HD above $z=1$.  Would it be better to measure one supernova (with $\sigma_{\mu}=0.30$ mag) at $z=1.7$ or one GRB (with $\sigma_{\mu}=0.65$) at $z=6.6$?  The model differences between the concordance and 'Riess' cosmologies is 0.15 mag at $z=1.7$ so the accuracy from a single supernova is twice the size of the effect being sought.  Alternatively, the model differences at $z=6.6$ is 1.00 mag, so the single high redshift GRB will detect the differences at close to the 1.5-sigma level.  In this case, the single GRB is three times better than the single maximal-redshift supernova.  That is, the long lever arm for GRBs more than makes up for their lower accuracy.  Or another way of looking at it is that a single GRB will provide more information than a single supernova.
	
	The above common argument is also wrong at a higher level.  This argument declares the $1.7<z \le 6.6$ region to be uninteresting within the concordance cosmology.  But this is presumptive that the concordance cosmology is correct.  (Similarly, back in 1997, I was told by a theorist that it was pointless to measure the HD to $z =1$ since it would merely confirm the then concordance cosmology of $\Omega_M=1$ and $\Omega_{\Lambda}=0$.)  Science advances by exploring unexplored regions and by performing critical tests of standard wisdom.  (And the standard wisdom of the now concordance cosmology is only a few years old.)  Who knows what we will find in the $1.7<z \le 6.6$ HD, and we won't know unless we look.  It would be unwise for the community to ignore $1.7<z \le 6.6$ as uninteresting, especially as the GRB data is {\it currently} flowing in for free from Swift.
	
	A disadvantage of supernovae for HD work is that the optical light can be dimmed with distance.   Two serious mechanisms have been proposed.  The first is that the optical light will be achromatically dimmed by grey dust and hence not corrected for in the usual dereddening (Aguirre 1999a; 1999b).  Simple versions of this mechanism have been excluded, but the possibility that the dust density changes with redshift has not been excluded (Riess et al. 2004).  The possibility of grey dust changing over time to match the predictions of the concordance cosmology seems contrived, but it is possible.  The second mechanism is that the optical light will be achromatically dimmed by refraction in Lyman-$\alpha$ clouds along the line of sight (Schild \& Dekker 2005).  This idea remains largely unexplored.  GRBs as cosmological tools are completely immune to both problems.

	A likely serious problem for the supernova HD is that the progenitor population of Type Ia supernova undoubtedly evolves in time, so that high redshift events might have a substantially different calibration from nearby events, thus leading to a distortion of the HD.  This is a reasonable possibility because the progenitors formed in the young universe will have lower metallicity than progenitors formed in recent times, and the metallicity might have a noticeable effect on the supernova peak brightnesses and decline rates.  The primary defense is that the spectra of supernovae at $z \lesssim 1$ appear similar to those of nearby events (Perlmutter et al. 1997; 1999), and this is good for putting some crude upper limit on the size of the evolution effects.  However, the peak brightnesses of supernovae are correlated with the galaxy type, and galaxy types evolve as we look back to higher redshifts, so we have observational evidence that the evolution effect is significant.  Detailed calculations under various scenarios (Dom\'{i}nguez, H\"{o}flich, \& Straniero 2001; 2003) show that the progenitor evolution effects are roughly 0.2 mag from $z=0$ to $z=1$.  This is comparable to the size of the cosmological effects over the same redshift range.  As such, the supernova HD cannot be used for precision cosmology until the evolution issue is resolved.  And there is currently no resolution of the long-standing question of whether the progenitors are recurrent novae, double white dwarfs, supersoft binary systems, or symbiotic stars.  Without knowing the identity of the progenitors, how can we evaluate the evolution of peak brightnesses as we go to high redshift?  In all, supernova cannot be applied to HD questions where systematic changes of less than $\sim$0.2 mag from low to high redshift are critical.
	
	Do GRBs have the same problem with evolution?  I argue that GRBs do {\it not} have any problems with evolution.  My reason is that the physical mechanisms that create the luminosity relations are thought to be based on light travel times, the degree of relativistic beaming, and energy conservation in the shocked material -- with none of these mechanisms changing as we look back in our universe.  The metallicity of the GRB progenitor or the surrounding ISM does not change the speed of light or the relativistic effects that are the basis for the luminosity relations.  The metallicity might affect the GRB luminosity, but then the physics of the luminosity relations would simply indicate the corresponding luminosity and the distance would be correctly deduced.  The population of GRBs might drift, for example, in luminosity as we look back to high redshift, but the luminosity indicators will still give the correct luminosity for each individual burst.  And this is all we need for GRBs to have zero evolution effects.  This argument should be examined further, but in the meantime, we are left with a situation where GRBs have arguably zero evolution effects whilst supernova have currently unknown evolution effects that can get as large as the signal being sought.
	
	Do either GRBs or Type Ia supernovae have an advantage concerning the number of outliers that must be rejected?  In Section 3, I rejected three GRBs and two rise times out of 72 bursts.  For comparison, Perlmutter et al. (1999) presented their main result with 6 rejected outliers out of 60 supernovae.  Riess et al. (2004) rejected 29 supernovae (out of 186) to form their 'gold sample', with the causes largely being due to uncertain classifications, too few points in the light curves, and high extinctions.  Many well-observed supernovae are known to be distant outliers (not even counting the super-luminous SN 1991T and sub-luminous SN 1991bg events); including S Andromedae (de Vaucouleurs \& Corwin 1985; Fesen, Hamilton, \& Saken 1989), SNLS-03D3bb (Howell et al. 2006), SN 2005hk (Phillips et al. 2006; Stanishev et al. 2006), SN 2002cx (Li et al. 2003), SN 2003gq (Jha et al. 2006), SN 2005P (Jha et al. 2006), SN 2005cc (Antilogus et al. 2005), and SN 1006 (Schaefer 1996a; 1996b).  Apparently, Type Ia supernovae have a higher percentage of outliers than GRBs, but I do not think that this is important for the relative merits.  Outliers are not important for either Type Ia supernovae or GRB HD work because the outliers can be readily identified and rejected for both types of explosions and because the outliers occupy a suitably small fraction of the overall event population.
	
	Supernovae have a substantial advantage for HD work because their physics is well known, whereas the physics of GRBs is relatively much less well known.  That is, the mechanisms of the supernova explosion have been very closely studied by many groups and we have excellent and extensive observational material.  In contrast, the basic scenario for GRBs (internal shocks within a relativistic jet produced by a core collapse of a very massive rotating star) is only a few years old, and many of the details are still being debated.  This situation leaves a comforting feeling when dealing with the supernova HD; while we are left to ponder whether there are any now-hidden surprises in the GRB HD.  This advantage for supernovae might only be psychological, however, because the calibrations and corrections in all supernova cosmology projects are entirely empirical.  That is, there is zero contribution from supernova theory in deriving the HD.  So nowhere has the advantage been used.  The GRB HD is also entirely empirical (see Figures 1-5), and thus it is operating on the same basis as the supernova HD.
	
	I have just made a strong case that the GRB HD has some good advantages over the supernova HD and that its disadvantages are not as bad as some might expect.  So which is 'better', or more to the point, which should get attention and resources?  For this, the answer is obvious that {\it both} should get attention and resources.  The reason is simply that no one method can convince the community by itself, so that multiple methods giving concordant results are required.  (This was a lesson learned deeply with the original supernova HD work in the late 1990s.)  That is, any answer from the GRB HD alone will not convince the community, just as any answer from SNAP alone will not convince the community.  What is needed is completely independent methods that concur.  To this end, our community must strongly support the SNAP mission, just as GRB HD efforts should be supported.
	
\section{Future Work}
	
	An obvious extension of this work is to perform simultaneous fits with GRBs and supernovae.  Care would have to be taken that each data set is handled correctly, for example with the GRB luminosity relations being fitted for every set of trial cosmological parameters.  The supernovae would dominate the delineation of the HD for $1<z$ while the GRBs would provide all the information for the HD for $1.7<z\le 6.6$.  This use of all data will provide the best possible constraints on the shape of the Hubble diagram.

	Another obvious extension of this work is to test a more versatile expansion for $w(z)$.  The $w(z)=w_0 + w\arcmin z$ expansion has problems with divergences for $z \gtrsim 1$.  The $w(z)=w_0 + w_a  z (1+z)^{-1}$ expansion makes a presumption that most of the change in $w$ occurs at $z \sim 1$ (which was about the only place where it could be tested with the supernovae) which might or might not be appropriate.  Perhaps an expansion like $w(z)=w_0 + w_a  z (z_{t}+z)^{-1}$, where $z_{t}$ is the redshift of the transition (Rappetti, Allen, \& Weller. 2005).  With the supernova plus GRB data covering such a wide range of redshift, it might be possible to go to additional parameters.

	I can think of a variety of places in which substantial improvements can be made within the next year or two.  The most obvious of these is that the luminosity relations can be improved.  To this end, Firmani et al. (2005; 2006) have found an improvement in the $E_{peak}-L$ relation by including duration information, and Li \& Paczy\'{n}ski (2006) have improved the $V_L$ relation with an alternative normalization factor and a different method of smoothing the light curve.  Other improvements can be expected, and let me list here a few ideas:  (1) We redshift and dilate the various luminosity indicators to the rest frame of the GRB with factors of $1+z$, but we measure the luminosity indicators and peak fluxes from light curves created for energy bands fixed in the Earth rest frame.  With Swift data and analysis software being public, it is now possible for investigators to calculate the luminosity indicators with light curves for standard energy bands in the rest frames of the individual GRBs.  (2) The proliferating definitions of variability all have many arbitrary choices, so I wonder if a completely different approach might result in a tighter relation.  One idea would be to look at the slope or intercept of the Fourier power spectrum, but I expect that this will be poor since the Fourier spectra are more appropriate for long data streams.  A better idea is to use a wavelet activity spectrum (as in Walker, Schaefer, \& Fenimore 2000) as wavelets are local measures and the activity directly measures the amplitude and time scale of the variations.  (3) Firmani et al. (2005; 2006) use a parameter called $T_{0.45}$, which is the duration during which the burst light curve is brighter than 45\% of the peak flux.  The choice of the 45\% level is only left over as a historical artifact and there is no real likelihood that this level is optimal.  An obvious study is to vary this level and seek the minimal scatter about the calibration equation.  And other duration definitions might be explored, for example $T_{90}$, $T_{90}/N_{peak}$, $T_{0.45}/N_{peak}$, or even $\tau_{RT}$.  (4) We now have many measurable parameters that are correlated with luminosity, and these might merely be projections of higher dimensional correlations.  It might be profitable to perform a multivariate analysis to try and discover some sort of a 'fundamental plane' which can be used to substantially reduce the scatter in the luminosity relations.
	
	Another substantial improvement in the near future will be that Swift (along with ground based observers) will be providing more bursts to include in the GRB HD.  The Konus-Wind and Suzaku-WAM experiments will provide tight limits on the $E_{peak}$ values for many bursts.  Swift has produced 22 GRBs for my sample in the year before the last entry.  In two years, Swift will have a total sample size of roughly 75 bursts.  One advantage of such a large Swift sample is that they will all come from a single instrument.  Such a homogeneity will produce more consistent luminosity relations and ease the calculations of Malmquist bias.  The sample over the next few years will also be important as an {\it independent} data set from the one in Table 2, and this will be a conceptually important means of testing the results from the current sample.

	It is possible to design a specialized space mission that will greatly improve the accuracy of the GRB HD.  For example, Lamb et al. (2005) describe a possible MIDEX-class mission that will measure $z$, $E_{peak}$, $S$, and $T_{jet}$ for $>$800 GRBs with $0.1<z<10$ during a two-year mission.  Such a program would measure $\Omega_M$ and $w_0$ to an accuracy of roughly $\pm 0.06$.
	
	The future will have the details of the GRB HD examined closely by many groups, with all the normal problems being raised and resolved.  With the GRB data now pouring in for free from Swift, our community will be seeing a lot of HD work over the near future.  But ultimately, any conclusions from the GRB HD will have to be tested and confirmed by independent methods.

\section{Conclusions}

	In this paper, I have constructed a Hubble diagram for 69 Gamma-Ray Bursts with redshifts from 0.17 to $>$6.  The basis for evaluating the luminosity distances is five luminosity indicators ($\tau _{lag}$, $V$, $E_{peak}$, $t_{jet}$, and $\tau _{RT}$).  The construction of the GRB HD has simply followed in the footsteps of many prior papers, and no new techniques are being used here.  What is new is that I am using over one order of magnitude more data than any previous work.  This is achieved by the simultaneous use of all five indicators (rather than only one or two at a time) as well as the use of recent bursts discovered by Swift.  My conclusions are:
	
	(1) The GRB Hubble diagram defines a well-behaved curve with half the bursts at $z>1.7$ and seven GRBs with $z>4$.  This bodes well on the utility and accuracy of cosmological information from the GRB HD.
	
	(2) The average one-sigma uncertainty for a single GRB is 0.65 mag in the distance modulus.  This is a factor of two larger than for the average one-sigma uncertainty for a single supernova.  This bodes well for the future, as there are a variety of means by which the GRB luminosity relations can likely be substantially improved.
	
	(3) The gravitational lensing and Malmquist biases are small, with the average bias of 0.03 mag and the RMS scatter of this bias being 0.14 mag.  This surprisingly small bias arises from both of two reasons; first that the average burst brightness (compared to the experimental threshold) varies by less than a factor of two as redshift goes from low to high, and second that the various competing effects (from increasing volume, increasing number density, decreasing luminosity function, and decreasing detection probability with increasing distances) all come close to canceling each other out for typical burst properties.
	
	(4) The $1.7<z\le 6.6$ region of the HD is of high interest and utility for two reasons (despite being matter dominated within the concordance model).  First, the high redshift provides a long lever arm so that slope changes will result in large differences in the distance modulus; for example a comparison between the concordance and 'Reiss' cosmologies at $z=1.7$ has only a difference of 0.15 mag while at $z=6.6$ the difference is 1.00 mag.  Second, it is good science to explore previously unexplored regions and to thoroughly test any recent concordance models, especially when the data is arriving even now and for free from Swift.
	
	(5) I argue that GRBs do not suffer from any evolution effects because the GRB luminosity indicators are based on light travel times, energy conservation, and the degree of relativistic beaming, none of which vary with metallicity or age of the universe.  The population of bursts might evolve in average luminosity, but the luminosity indicators will still yield the correct luminosity of each individual burst.
	
	(6) I find that the current GRB HD is consistent with the concordance cosmology of $w=-1$, $\Omega_M=0.27$, and $\Omega_{\Lambda}=0.73$.  That is, there is no evidence for the Cosmological Constant being inconstant.

\acknowledgements
	This work is only possible from the results of many workers over the past decade and more who have spent vast times and efforts accumulating the observations summarized in the data tables.  These includes the designers, builders, and analysts for the burst detectors onboard the  GRO, Konus, SAX, HETE-2, Integral, and Swift satellites as well as the many optical observers who have sleepless nights to discover the optical transients, measure their light curves, and get their redshifts.  The key task of this enterprise is the fast distribution of burst information via the GCN by Scott Barthelmy. 
	
	I thank Hans Krimm for providing many spectral fit parameters from Swift data, Sergey Golenetskii for passing along various $E_{peak}$ and $P$ values from the Konus-Wind experiment, Sheila McBreen for providing the INTEGRAL light curve and fit values for GRB 050502, and Hugo Martel for sending numerical results of his gravitational lensing probability distributions.  I also thank Jochen Greiner for maintaining his web page with an exhaustive collection of the data in a convenient format.  Neil Gehrels, Hans Krimm, and David Palmer helped in improving the manuscript for this paper.
	
	This work is supported in part by NASA under grant NNG06GH07G.

\clearpage

\begin{deluxetable}{llll}
\tabletypesize{\scriptsize}
\tablecaption{Prior GRB Hubble Diagrams.
\label{tbl1}}
\tablewidth{0pt}
\tablehead{
\colhead{Reference}   &
\colhead{GRBs}   &
\colhead{Relations Used\tablenotemark{a}}   &
\colhead{Number of Indicators} 
}
\startdata
Schaefer (2003a) & 9 & $\tau _{lag}$, $V$ & 18 \\
Bloom, Frail, \& Kulkarni (2003) & 16\tablenotemark{b} & $E _{\gamma}$ & 16 \\
Xu, Dai, \& Liang (2005) & 17 & $t _{jet}$ & 17 \\
Firmani et al. (2005) & 15 & $t _{jet}$ & 15 \\
Liang \& Zhang (2005) & 15 & $t _{jet}$ & 15 \\
Firmani et al. (2006) & 19 & $T_{0.45}$ & 19 \\
Schaefer (this work) & 69 & $\tau _{lag}$, $V$, $E_{peak}$, $t_{jet}$, $\tau _{RT}$ & 243 \\
\enddata
    
\tablenotetext{a}{The first five relations in section 3 are indicated by $\tau _{lag}$, $V$, $E_{peak}$, $t_{jet}$, and $\tau _{RT}$ respectively.  The $E_{\gamma}$ indicates the $t_{jet}$ relation but without the later improvement of the correlation with $E_{peak}$  The $T_{0.45}$ (a measure of the burst duration) indicates a relation between the product of powers of $E_{peak}$ and $T_{0.45}$ and the isotropic luminosity.}

\tablenotetext{b}{An additional 13 GRBs are discussed but these either do not have redshifts or have only limits on the jet break time.}

\end{deluxetable}

\clearpage

\begin{deluxetable}{lllllllllll}
\tabletypesize{\scriptsize}
\rotate
\tablecaption{Burst redshift, peak flux, and spectral data.
\label{tbl2}}
\tablewidth{0pt}
\tablehead{
\colhead{GRB}   &
\colhead{Experiment}   &
\colhead{$z$}  &
\colhead{Ref.\tablenotemark{a}}  &
\colhead{$P\tablenotemark{b}\tablenotemark{c}$}   &
\colhead{$E_{min}-E_{max}$ (keV)}   &
\colhead{Ref.\tablenotemark{a}}   &
\colhead{$E_{peak}$ (keV){\tablenotemark{c}}} &
\colhead{$\alpha \tablenotemark{d}$}   &
\colhead{$\beta \tablenotemark{d}$}   &
\colhead{Ref.\tablenotemark{a}}
}
\startdata
970228	&	Konus	&	0.70	&	1	&	3.7E-6	$\pm$	1E-7	&	40	-	700	&	2	&	115	 -	38	+	38	&	-1.54	$\pm$	0.08	&	-2.5	$\pm$	0.4	&	2	\\
970508	&	BATSE	&	0.84	&	3	&	7.4E-7	$\pm$	[7E-6]	&	50	-	300	&	4	&	389	 -	[40]	+	[40]	&	-1.19	$\pm$	[0.1]	&	-1.83	$\pm$	[0.4]	&	4	\\
970828	&	Konus	&	0.96	&	5	&	3.0E-6	$\pm$	[3E-7]	&	50	-	300	&	4	&	298	 -	[30]	+	[30]	&	-0.704	$\pm$	[0.1]	&	-2.072	$\pm$	[0.4]	&	4	\\
971214	&	BATSE	&	3.42	&	6	&	1.95	$\pm$	0.05	&	50	-	300	&	7	&	190	 -	[20]	+	[20]	&	-0.78	$\pm$	[0.1]	&	-2.57	$\pm$	[0.4]	&	4	\\
980613	&	SAX	&	1.10	&	8	&	1.6E-7	$\pm$	4E-8	&	40	-	700	&	2	&	92	 -	42	+	42	&	-1.43	$\pm$	0.24	&	-2.7	$\pm$	0.6	&	2	\\
980703	&	BATSE	&	0.97	&	9	&	2.40	$\pm$	0.06	&	50	-	300	&	7	&	254	 -	[25]	+	[25]	&	-1.31	$\pm$	[0.1]	&	-2.4	$\pm$	[0.4]	&	4	\\
990123	&	BATSE	&	1.61	&	10	&	16.41	$\pm$	0.12	&	50	-	300	&	7	&	604	 -	[60]	+	[60]	&	-0.9	$\pm$	[0.1]	&	-2.48	$\pm$	[0.4]	&	4	\\
990506	&	BATSE	&	1.31	&	11	&	18.56	$\pm$	0.13	&	50	-	300	&	7	&	283	 -	[30]	+	[30]	&	-1.37	$\pm$	[0.1]	&	-2.15	$\pm$	[0.4]	&	4	\\
\enddata
\end{deluxetable}

\clearpage

\begin{deluxetable}{llllllllll}
\tabletypesize{\scriptsize}
\rotate
\tablecaption{Burst fluence and beaming data.
\label{tbl3}}
\tablewidth{0pt}
\tablehead{
\colhead{$GRB$}   &
\colhead{$S$ (erg cm$^{-2}$)\tablenotemark{a}}   &
\colhead{$E_{min}-E_{max}$ (keV)}   &
\colhead{Ref. \tablenotemark{b}}   &
\colhead{$S_{bolo}$ (erg cm$^{-2}$)} &
\colhead{$t_{jet}$ (days)}   &
\colhead{Ref. \tablenotemark{b}}  &
\colhead{$\theta_{jet}$ (deg.)\tablenotemark{a}}   &
\colhead{Ref. \tablenotemark{b}} &
\colhead{$F_{beam}$} 
}
\startdata
970508	&	5.5E-6	$\pm$	[5E-7]	&	20	 --	2000	&	1	&	8.09E-6	$\pm$	8.1E-7	&	25	$\pm$	5	&	2	&	23	$\pm$	3	&	3	&	0.0795	$\pm$	0.0204	\\
970828	&	9.6E-5	$\pm$	[1E-5]	&	20	 --	2000	&	1	&	1.23E-4	$\pm$	1.2E-5	&	2.2	$\pm$	0.4	&	2	&	5.91	$\pm$	0.79	&	4	&	0.0053	$\pm$	0.0014	\\
980703	&	2.3E-5	$\pm$	[2E-6]	&	20	 --	2000	&	1	&	2.83E-5	$\pm$	2.9E-6	&	3.4	$\pm$	0.5	&	2	&	11.02	$\pm$	0.8	&	4	&	0.0184	$\pm$	0.0027	\\
990123	&	2.7E-4	$\pm$	[3E-5]	&	20	 --	2000	&	1	&	3.11E-4	$\pm$	3.1E-5	&	2.04	$\pm$	0.46	&	5	&	3.98	$\pm$	0.57	&	4	&	0.0024	$\pm$	0.0007	\\
990510	&	2.3E-5	$\pm$	[2E-6]	&	20	 --	2000	&	1	&	2.85E-5	$\pm$	2.9E-6	&	1.6	$\pm$	0.2	&	6	&	3.74	$\pm$	0.28	&	4	&	0.0021	$\pm$	0.0003	\\
990705	&	7.5E-5	$\pm$	8E-6	&	40	 --	700	&	7	&	1.34E-4	$\pm$	1.5E-5	&	1	$\pm$	0.2	&	2	&	4.78	$\pm$	0.66	&	4	&	0.0035	$\pm$	0.0010	\\
990712	&	6.5E-6	$\pm$	3E-7	&	40	 --	700	&	7	&	1.19E-5	$\pm$	6.2E-7	&	1.6	$\pm$	0.2	&	8	&	9.47	$\pm$	1.2	&	4	&	0.0136	$\pm$	0.0034	\\
991216	&	1.9E-4	$\pm$	[2E-5]	&	20	 --	2000	&	1	&	2.48E-4	$\pm$	2.5E-5	&	1.2	$\pm$	0.4	&	2	&	4.44	$\pm$	0.7	&	4	&	0.0030	$\pm$	0.0009	\\
\enddata
 
\tablenotetext{a}{The uncertainties given in square brackets are conservative estimates for when no error bar is quoted in the original literature.}
   
\tablenotetext{b}{References. --- (	1	)	Jimenez, Band, \& Piran 2001;
(	2	)	Bloom et al. 2003b;
(	3	)	Ghirlanda et al. 2004;
(	4	)	Nava et al. 2006;
(	5	)	Kulkarni et al. 1999;
(	6	)	Israel et al. 1999;
(	7	)	Amati et al. 2002;
(	8	)	Bjornsson et al. 2001;
(	11	)	Frontera et al. 2001;
(	12	)	Jakobsson et al. 2003;
(	13	)	Sakamoto et al. 2005a;
(	14	)	Berger et al. 2003;
(	15	)	Price et al. 2003a;
(	16	)	Holland et al. 2003;
(	17	)	Klose et al. 2004;
(	18	)	Andersen et al. 2003;
(	19	)	Jakobsson et al. 2004;
(	20	)	Galassi et al. 2004;
(	21	)	Stanek et al. 2005;
(	22	)	Krimm et al. 2005;
(	23	)	Perri et al. 2005;
(	24	)	Hullinger et al. 2005;
(	25	)	Hurkett et al. 2006;
(	26	)	Cummings et al. 2005;
(	27	)	Blustin et al. 2006;
(	28	)	Sakamoto et al. 2006a;
(	29	)	Tagliaferri et al. 2005;
(	30	)	Frail et al. 2006;
(	33	)	This work
(	34	)	Golenetskii et al. 2005;
(	35	)	Racusin et al. 2005;
(	36	)	Golenetskii et al. 2006;
(	37	)	Romano et al. 2006;
(	38	)	Sakamoto et al. 2006d;
(	39	)	Dai \& Stanek 2006;
(	40	)	Markwardt et al. 2006;
(	41	)	Moretti; et al. 2006;
}

\end{deluxetable}

\clearpage
\begin{deluxetable}{llllllllll}
\tabletypesize{\scriptsize}
\rotate
\tablecaption{Luminosities and Luminosity Indicators.
\label{tbl4}}
\tablewidth{0pt}
\tablehead{
\colhead{$GRB$}   &
\colhead{$z$}   &
\colhead{$P_{bolo}$ (erg cm$^{-2}$ s$^{-1}$)}   &
\colhead{$S_{bolo}$ (erg cm$^{-2}$)} &
\colhead{$F_{beam}$} &
\colhead{$\tau_{lag}$ (sec)}   &
\colhead{$V$}  &
\colhead{$E_{peak}$ (keV)\tablenotemark{a}}   &
\colhead{$\tau_{RT}$ (sec)} &
\colhead{$N_{peak}$}
}
\startdata
970228	&	0.70	&	7.3E-6	$\pm$	4.3E-7	&		É		&		É		&		É		&	0.0059	$\pm$	0.0008	&	115	 -	38	+	38	&	0.26	$\pm$	0.04	&	2	\\
970508	&	0.84	&	3.3E-6	$\pm$	3.3E-7	&	8.09E-6	$\pm$	8.1E-7	&	0.0795	$\pm$	0.0204	&	0.5	$\pm$	0.3	&	0.0047	$\pm$	0.0009	&	389	 -	[40]	+	[40]	&	0.71	$\pm$	0.06	&	1	\\
970828	&	0.96	&	1.0E-5	$\pm$	1.1E-6	&	1.23E-4	$\pm$	1.2E-5	&	0.0053	$\pm$	0.0014	&		É		&	0.0077	$\pm$	0.0007	&	298	 -	[30]	+	[30]	&	0.26	$\pm$	0.07	&	8	\\
971214	&	3.42	&	7.5E-7	$\pm$	2.4E-8	&		É		&		É		&	0.03	$\pm$	0.03	&	0.0153	$\pm$	0.0006	&	190	 -	[20]	+	[20]	&	0.05	$\pm$	0.02	&	12	\\
980613	&	1.10	&	3.0E-7	$\pm$	8.3E-8	&		É		&		É		&		É		&		É		&	92	 -	42	+	42	&		É		&	É	\\
980703	&	0.97	&	1.2E-6	$\pm$	3.6E-8	&	2.83E-5	$\pm$	2.9E-6	&	0.0184	$\pm$	0.0027	&	0.4	$\pm$	0.1	&	0.0064	$\pm$	0.0003	&	254	 -	[25]	+	[25]	&	3.60	$\pm$	0.5	&	1	\\
990123	&	1.61	&	1.3E-5	$\pm$	5.0E-7	&	3.11E-4	$\pm$	3.1E-5	&	0.0024	$\pm$	0.0007	&	0.16	$\pm$	0.03	&	0.0175	$\pm$	0.0001	&	604	 -	[60]	+	[60]	&		É		&	3	\\
990506	&	1.31	&	1.1E-5	$\pm$	1.5E-7	&		É		&		É		&	0.04	$\pm$	0.02	&	0.0131	$\pm$	0.0001	&	283	 -	[30]	+	[30]	&	0.17	$\pm$	0.03	&	12	\\
\enddata
 
\tablenotetext{a}{The uncertainties given in square brackets are conservative estimates for when no error bar is quoted in the original literature.}

\end{deluxetable}
\clearpage

\begin{deluxetable}{llll}
\tabletypesize{\scriptsize}
\tablecaption{Light Curve Properties.
\label{tbl5}}
\tablewidth{0pt}
\tablehead{
\colhead{Experiment}   &
\colhead{For $V$, $\tau_{RT}$, $N_{peak}$}   &
\colhead{For $\tau_{lag}$} &
\colhead{Time Resolution (sec)}
}
\startdata
BATSE & 25-320 keV & 25-50 and 100-320 keV & 0.064 \\
Konus & 50-200 keV & only one channel available & 0.064 \\
HETE-2 & 30-400 keV & 7-40 and 30-400 keV & 0.164 \\
Swift & 25-350 keV & 25-50 and 100-350 keV & 0.064  \\
\enddata

\end{deluxetable}

\clearpage

\begin{deluxetable}{lllllllll}
\tabletypesize{\scriptsize}
\tablecaption{Derived Distance Moduli.
\label{tbl6_01}}
\tablewidth{0pt}
\tablehead{
\colhead{$GRB$}   &
\colhead{$z$}   &
\colhead{$\mu_1\tablenotemark{a}$}   &
\colhead{$\mu_2\tablenotemark{a}$} &
\colhead{$\mu_3\tablenotemark{a}$} &
\colhead{$\mu_4\tablenotemark{a}$}   &
\colhead{$\mu_5\tablenotemark{a}$}  &
\colhead{$\mu\tablenotemark{a}$}   &
\colhead{$\mu\tablenotemark{b}$}
}
\startdata
970228	&	0.70	&		É		&	42.53	$\pm$	1.31	&	42.41	$\pm$	1.16	&		É		&	43.42	$\pm$	1.21	&	42.79	$\pm$	0.71	&	42.56	$\pm$	0.70	\\
970508	&	0.84	&	43.09	$\pm$	1.19	&	43.13	$\pm$	1.34	&	45.66	$\pm$	1.04	&	43.92	$\pm$	0.63	&	43.09	$\pm$	1.20	&	43.92	$\pm$	0.43	&	43.68	$\pm$	0.44	\\
970828	&	0.96	&		É		&	42.99	$\pm$	1.28	&	44.07	$\pm$	1.03	&	43.61	$\pm$	0.63	&	43.27	$\pm$	1.25	&	43.58	$\pm$	0.46	&	43.32	$\pm$	0.48	\\
971214	&	3.42	&	48.76	$\pm$	1.50	&	48.69	$\pm$	1.19	&	47.55	$\pm$	1.04	&		É		&	49.33	$\pm$	1.33	&	48.44	$\pm$	0.62	&	48.06	$\pm$	0.61	\\
980613	&	1.10	&		É		&		É		&	45.85	$\pm$	1.33	&		É		&		É		&	45.85	$\pm$	1.33	&	45.63	$\pm$	1.30	\\
980703	&	0.97	&	44.47	$\pm$	1.03	&	44.91	$\pm$	1.28	&	46.06	$\pm$	1.02	&	43.59	$\pm$	0.58	&	42.11	$\pm$	1.20	&	44.07	$\pm$	0.40	&	43.91	$\pm$	0.42	\\
990123	&	1.61	&	43.23	$\pm$	1.02	&	44.83	$\pm$	1.21	&	45.59	$\pm$	1.05	&	45.54	$\pm$	0.66	&		É		&	44.99	$\pm$	0.45	&	44.63	$\pm$	0.47	\\
990506	&	1.31	&	44.82	$\pm$	1.15	&	44.24	$\pm$	1.23	&	44.19	$\pm$	1.03	&		É		&	43.96	$\pm$	1.22	&	44.31	$\pm$	0.57	&	43.99	$\pm$	0.57	\\
\enddata

\tablenotetext{a}{This assumes the concordance cosmology with $\Omega_M=0.27$ and $w=-1$.  The units of the distance moduli are magnitudes.}

\tablenotetext{b}{This assumes the best fit cosmology from the gold sample of Riess et al. (2004) with $w_0 = -1.31$ and $w\arcmin = 1.48$.}

\end{deluxetable}

\clearpage

\begin{deluxetable}{llllll}
\tabletypesize{\scriptsize}
\tablecaption{Gravitational Lensing and Malmquist Biases.
\label{tbl6_02}}
\tablewidth{0pt}
\tablehead{
\colhead{$GRB$}   &
\colhead{Experiment}   &
\colhead{$z$}   &
\colhead{$L$ (erg s$^{-1}$}  &
\colhead{$P_{bolo}$ (erg cm$^{-2}$ s$^{-1}$)}   &
\colhead{$\mu_{bias}$ (mag)}
}
\startdata
970228	&	Konus	&	0.695	&	1.7E+52	$\pm$	4.0E+51	&	7.3E-6	$\pm$	4.3E-7	&	0.05	$\pm$	0.01	\\
970508	&	BATSE	&	0.835	&	6.5E+51	$\pm$	9.5E+50	&	3.3E-6	$\pm$	3.3E-7	&	0.01	$\pm$	0.01	\\
970828	&	Konus	&	0.9578	&	8.5E+51	$\pm$	1.8E+51	&	1.0E-5	$\pm$	1.1E-6	&	0.05	$\pm$	0.01	\\
971214	&	BATSE	&	3.418	&	2.1E+53	$\pm$	6.4E+52	&	7.5E-7	$\pm$	2.4E-8	&	-0.15	$\pm$	0.02	\\
980613	&	SAX	&	1.0964	&	8.1E+51	$\pm$	7.0E+51	&	3.0E-7	$\pm$	8.3E-8	&	0.08	$\pm$	0.03	\\
980703	&	BATSE	&	0.966	&	2.7E+51	$\pm$	3.9E+50	&	1.2E-6	$\pm$	3.6E-8	&	-0.01	$\pm$	0.01	\\
990123	&	BATSE	&	1.61	&	3.0E+52	$\pm$	5.7E+51	&	1.3E-5	$\pm$	5.0E-7	&	0.06	$\pm$	0.01	\\
990506	&	BATSE	&	1.30658	&	6.1E+52	$\pm$	1.3E+52	&	1.1E-5	$\pm$	1.5E-7	&	0.07	$\pm$	0.01	\\
\enddata
\end{deluxetable}

\clearpage

\begin{deluxetable}{lllllll}
\tabletypesize{\scriptsize}
\tablecaption{How Robust are the $\chi^2$ Contours?
\label{tbl8}}
\tablewidth{0pt}
\tablehead{
\colhead{Case}   &
\colhead{GRBs}   &
\colhead{Best $\chi^2$} &
\colhead{Position of long axis} &
\colhead{Constraint on $w_0$} &
\colhead{$\chi^2$($w=-1$)} &
\colhead{$\chi^2$('Riess')}
}
\startdata
Standard	&	69	&	70.7	&	$w_a + 2.1w_0 = -1.0 \pm 0.3$	&	$w_0 = 0.2 \pm 0.6$	&	72.3	&	80.3	 \\
Minus 2 highest-$z$ GRBs	&	67	&	68.4	&	$w_a + 2.1w_0 = -0.8 \pm 0.3$	&	$w_0 = 0.5 \pm 2$	&	69.5	&	73.0	 \\
Plus 5 outliers	&	72	&	209.9	&	$w_a + 2.2w_0 = -0.6 \pm 0.3$	&	$w_0 > 2$	&	240.9	&	247.0	 \\
Plus 4 outliers (without GRB 980425)	&	71	&	96.7	&	$w_a + 2.2w_0 = -0.4 \pm 0.3$	&	$w_0 > 0$	&	105.1	&	110.2	 \\
Minus GRBs 050802, 060108, 060108	&	66	&	67.4	&	$w_a + 2.1w_0 = -0.8 \pm 0.3$	&	$w_0 = 0.6 \pm 0.9$	&	69.6	&	77.7	 \\
Minus $\tau_{lag}-L$ relation	&	69	&	67.8	&	$w_a + 2.1w_0 = -1.4 \pm 0.4$	&	$w_0 = -0.3 \pm 1.2$	&	67.9	&	76.7	 \\
Minus $V-L$ relation	&	69	&	58.3	&	$w_a + 2.1w_0 = -0.6 \pm 0.3$	&	$w_0 = 0.2 \pm 1.4$	&	60.3	&	64.6	 \\
Minus $E_{peak}-L$ relation	&	67	&	86.0	&	$w_a + 2.2w_0 = -0.4 \pm 0.3$	&	$w_0 = 0.4 \pm 1.3$	&	88.3	&	90.7	 \\
\enddata

\end{deluxetable}

\clearpage

\begin{figure}
\epsscale{.80}
\plotone{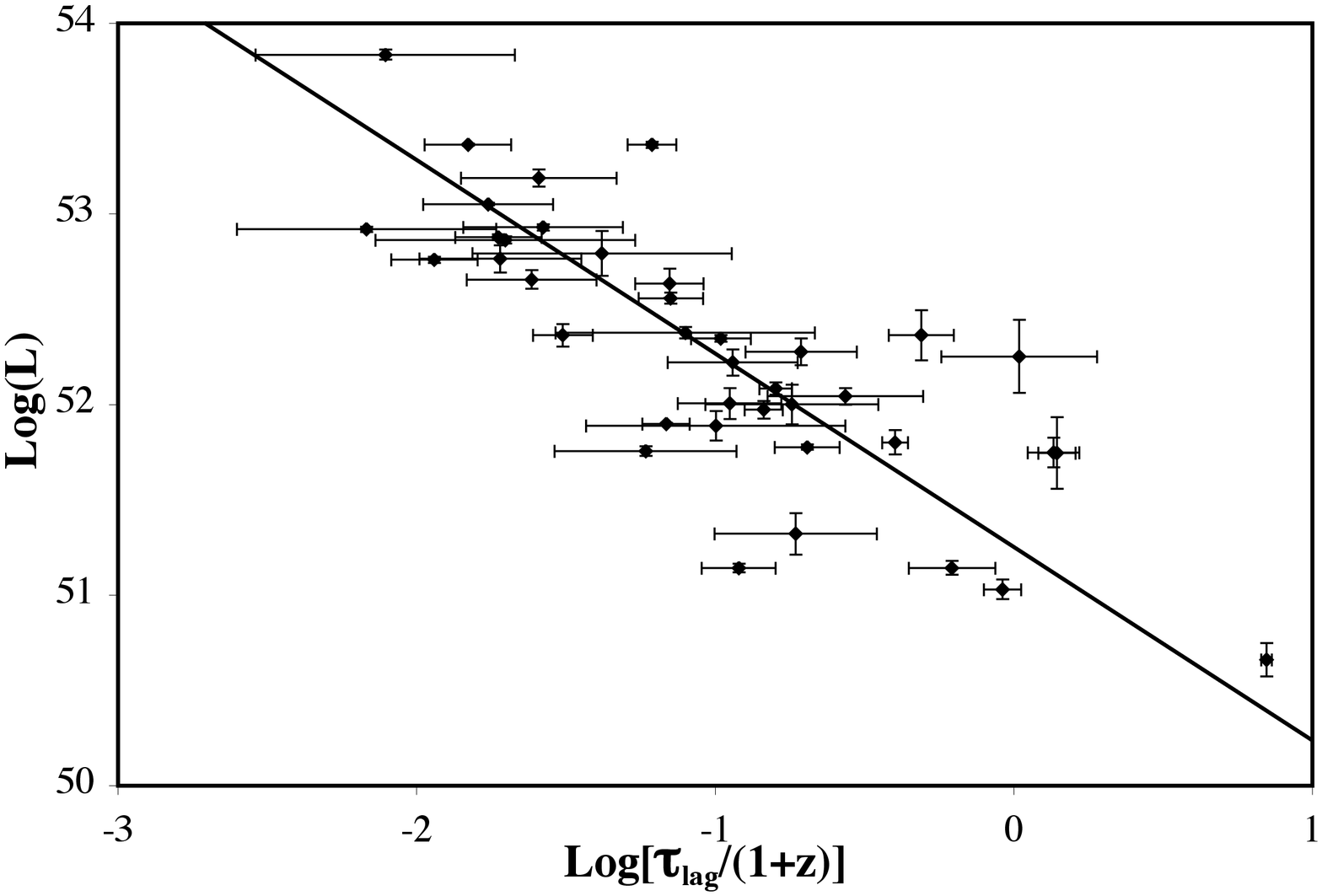}
\caption{
Lag -- luminosity relation.  Bursts with long lags are low luminosity, while bursts with short lags are high luminosity.  The lag times for 39 GRBs have been corrected to the rest frame of the GRB and plotted versus the isotropic luminosity with the best fit power law (see equation 12) superposed.  The one-sigma measurement uncertainties were used for the error bars.  The lag-luminosity relation is thought to be caused by the speed of the radiative cooling of the shocked material (and hence the lag) being determined by the luminosity (Schaefer 2004).  The observed slope in this Figure (-1.01) is close to the theoretically predicted slope of -1.  The point of this Figure is that the lag time can be measured from $\gamma$-ray data alone and then a luminosity distance and a luminosity can be deduced for all GRBs.}

\end{figure}

\clearpage

\begin{figure}
\epsscale{.80}
\plotone{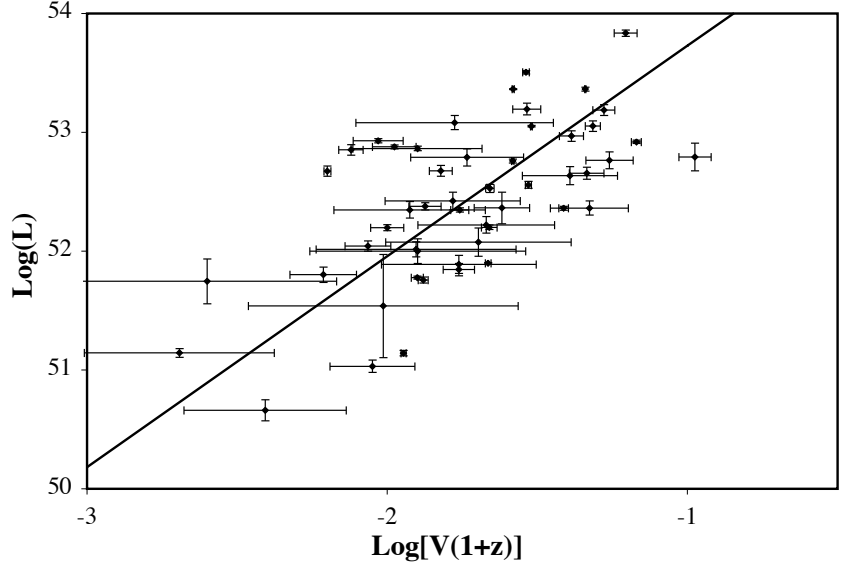}
\caption{
Variability -- luminosity relation.  The variabilities for 51 GRBs have been corrected to the rest frame of the GRB and plotted versus the isotropic luminosity with the best fit power law (see equation 14) superposed.  The one-sigma measurement uncertainties were used for the error bars.  The variability-luminosity relation is thought to be caused by the fact that both the variability and luminosity are functions of the bulk Lorentz factor of the jet ($\Gamma_{jet}$).  The variability is a measure of the sharpness of the pulse structure, which is determined by the size of the visible region in the jet which is determined by $\Gamma_{jet}$.  There is substantial scatter in this relation, so the variability carries less distance information than does the lag or $E_{peak}$.  For low luminosity GRBs, the variability becomes small so that it is comparable or smaller than the ordinary Poisson noise in the light curve and hence the uncertainties in $V$ become large.}

\end{figure}

\clearpage

\begin{figure}
\epsscale{.80}
\plotone{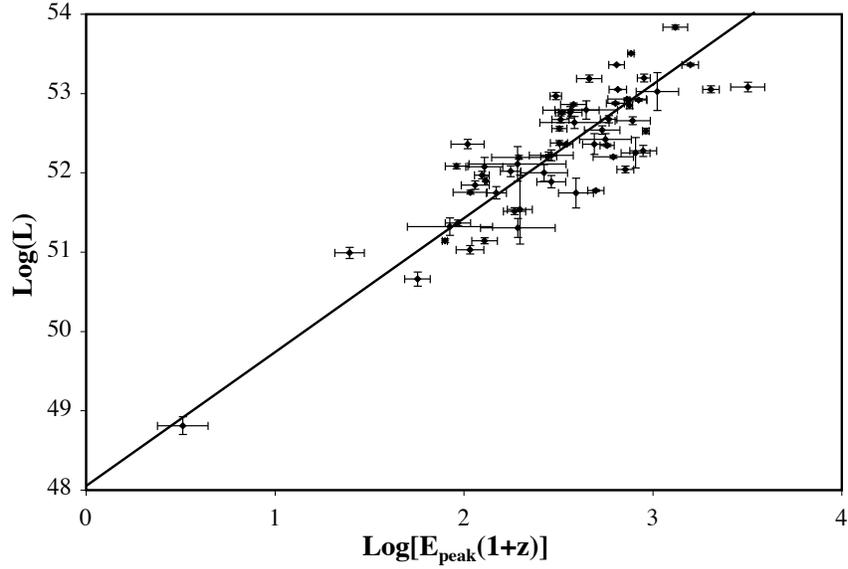}
\caption{
$E_{peak}$ -- luminosity relation.  As $\Gamma_{jet}$ for a burst increases, the luminosity ($L$) increases as a power law of $\Gamma_{jet}$ while the $E_{peak}$ also increases proportionally to $\Gamma_{jet}$ due to the blue shifting of the emitting region (Schaefer 2003b).  This prediction that $E_{peak}$ would be a luminosity indicator is confirmed by this figure.   The $E_{peak}$ values for 64 GRBs have been corrected to the rest frame of the GRB and plotted versus the isotropic luminosity with the best fit power law (see equation 16) superposed.  This relation is {\it not} the 'Amati relation' (the $E_{peak}$--$E_{\gamma,iso}$ relation; Amati et al. 2002), which is based on different physics (similar to the $E_{peak}$--$E_{\gamma}$ relation).}

\end{figure}

\clearpage

\begin{figure}
\epsscale{.80}
\plotone{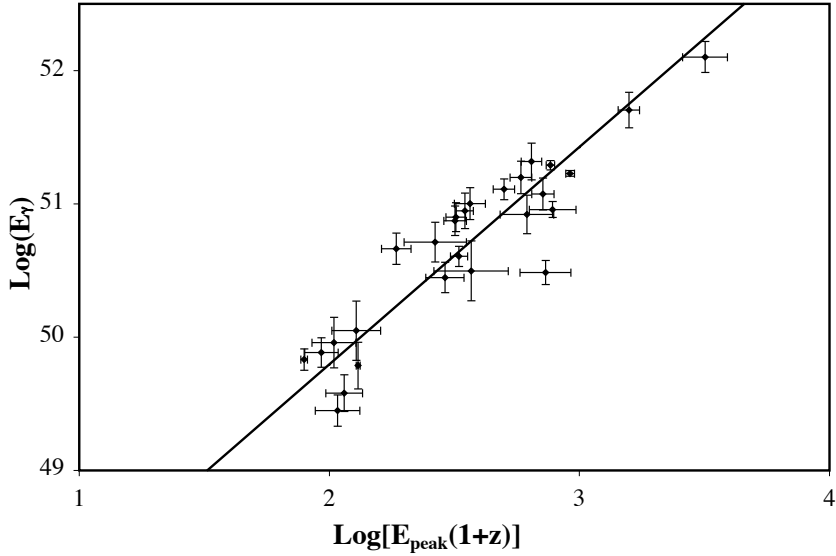}
\caption{
$E_{peak}$ -- $E_{\gamma}$ relation.  The $E_{peak}$ values for 27 GRBs have been corrected to the rest frame of the GRB and plotted versus the total burst energy in the $\gamma$-rays ($E_{\gamma}$) with the best fit power law (see equation 18) superposed.  This relation is the tightest of the GRB luminosity indicators, and so the derived distance will carry higher weight than distances from, say, the $V$--$L$ relation.  To be included in this plot, the GRB afterglow must have an observed jet break in its light curve, and this means that only a fraction of GRBs with redshifts can yield a distance measure from this relation.  This relation has been used by various groups (e.g., Ghirlanda, Ghisellini, \& Lazzati 2004; Xu, Dai, \& Liang 2005) but never with more than 18 GRBs included (Nava et al. 2006).  This paper includes 9 more events (a 50\% increase in the sample) mainly due to new bursts from Swift.}

\end{figure}

\clearpage

\begin{figure}
\epsscale{.80}
\plotone{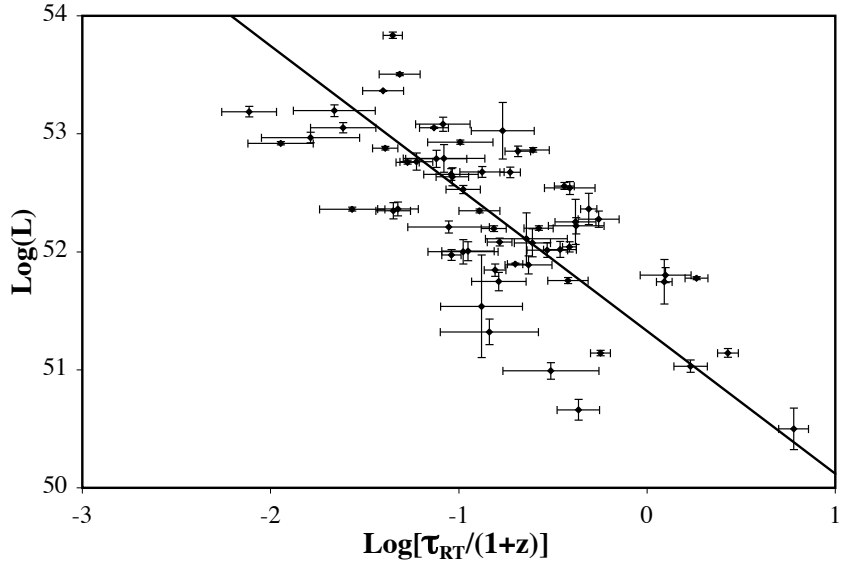}
\caption{
Minimum rise time -- luminosity relation.  This luminosity relation was originally predicted because the luminosity depends as a power law of the bulk Lorentz factor of the jet ($\Gamma_{jet}$) while the minimum rise time in the GRB light curve also depends on $\Gamma_{jet}$ through the size of the visible region in the shocked jet (Schaefer 2002).  This Figure provides a test and confirmation of the prediction that $\tau_{RT}$ is a correlated with luminosity as a power law.  The rise times for 62 GRBs have been corrected to the rest frame of the GRB and plotted versus the isotropic luminosity with the best fit power law (see equation 20) superposed.  The one-sigma measurement uncertainties were used for the error bars.  }

\end{figure}

\clearpage

\begin{figure}
\epsscale{.80}
\plotone{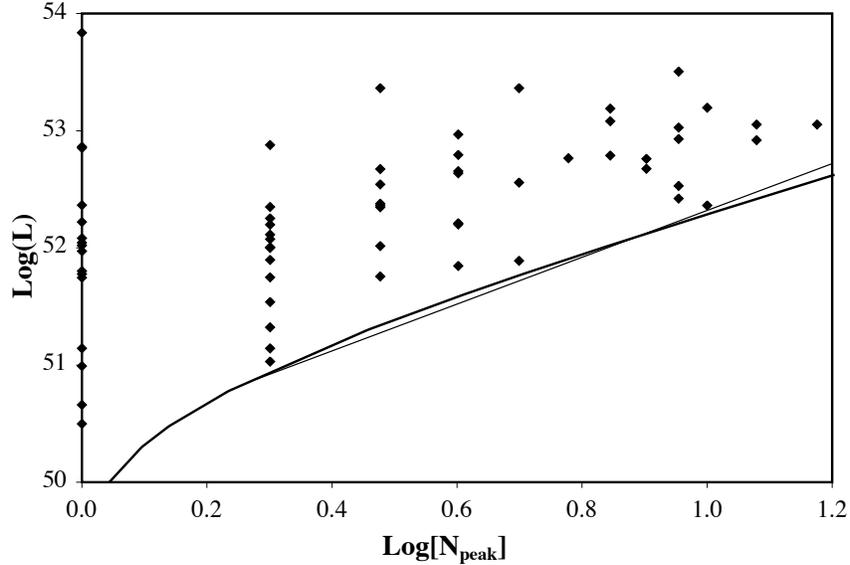}
\caption{
$N_{peak}$ -- luminosity relation.  With pulse rise times and durations depending on the opening angle of the visible region (and hence on $\Gamma_{jet}$), low luminosity bursts (with low $\Gamma_{jet}$) will have very broad pulses which inevitably blur together no matter how many pulses are in the light curve.  But a high luminosity event (with high $\Gamma_{jet}$) will have short pulse durations so that pulses are unlikely to be overlapping.  Thus, a theoretical prediction (Schaefer 2002) is that a lower limit on the burst luminosity can be derived from a count of the number of peaks in the light curve.  This Figure shows a test and confirmation of this prediction, as all 64 GRBs lie above the theoretical limit (the curved line).  To a close approximation for two-or-more peaks, the limit on luminosity is a power law (see equation 22) as displayed by a narrow line segment in the Figure.  This relation only provides a lower limit on the luminosity, and thus is not of any use for constructing a GRB Hubble diagram.  But this relation is useful for quickly putting limits on redshifts as an alert to observers, for example a faint Swift burst with many peaks is rapidly known to have high redshift.}

\end{figure}

\clearpage

\begin{figure}
\epsscale{1.1}
\plotone{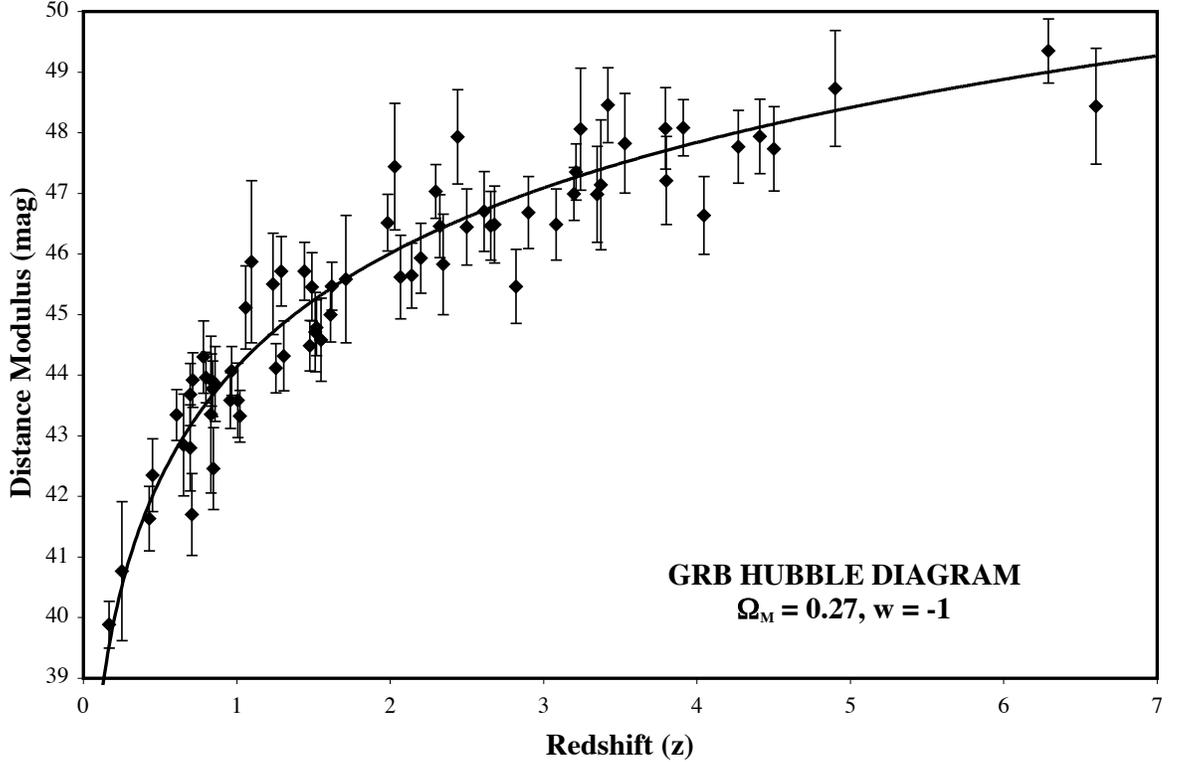}
\caption{
GRB Hubble diagram for the concordance cosmology.  For the concordance cosmology (for a flat universe with $\Omega_{M}=0.27$ and $w=-1$), this HD contains 69 GRBs out to redshift of $>$6.  On this plot, look to see the redshift regime covered by supernova, with almost all being with $z<1$ and only 10 with $1<z\lesssim 1.7$.  The curve is the model as given by equation 10.  We see that the observations follow smoothly along the model, indeed with chi-square of 72.3 and a reduced chi-square of 1.05.  This is to say that the GRB HD is consistent with an unchanging Dark Energy.}

\end{figure}

\clearpage

\begin{figure}
\epsscale{1.1}
\plotone{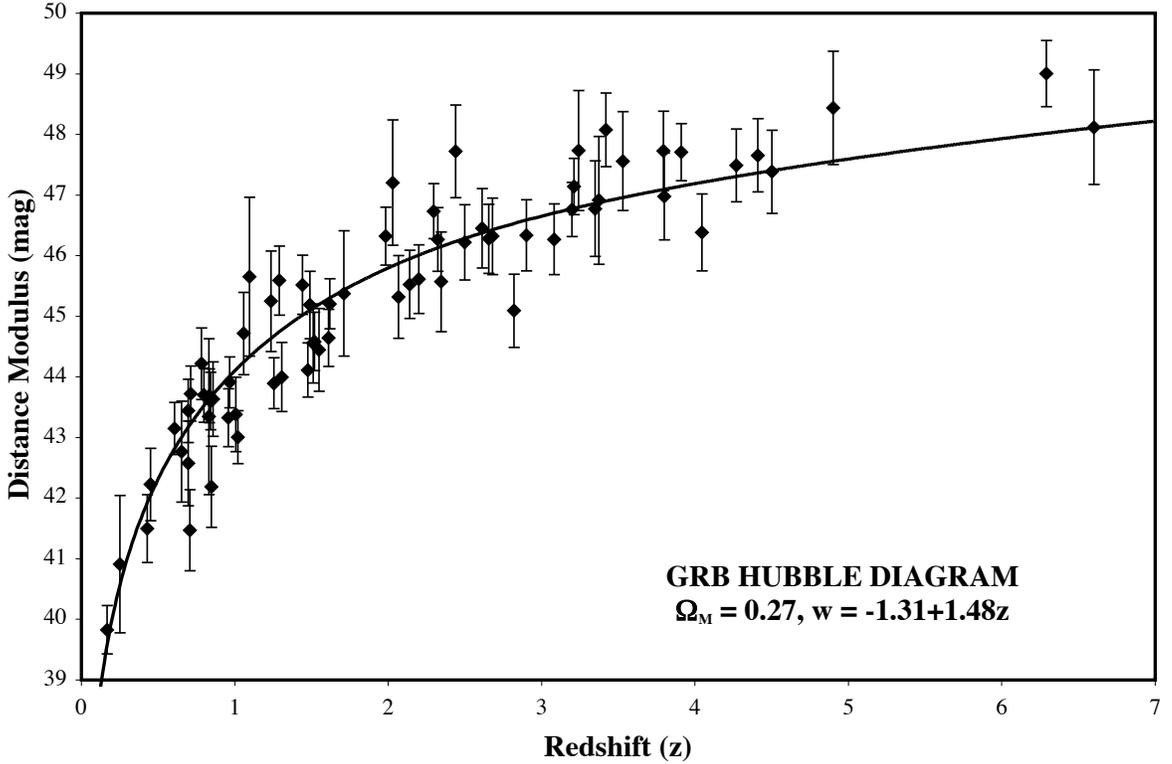}
\caption{
GRB Hubble diagram for Riess' best fit cosmology.  Riess et al. (2004) fitted a large sample of supernovae including those discovered by the Hubble Space Telescope (the so-called 'gold sample') and found that their best fit cosmology had the Dark Energy changing with time as described by $w_0 = -1.31$ and $w\arcmin = 1.48$.  With this, the equation of state parameter for the Dark Energy ($w$) varies with redshift as $w=-1.31+1.48z$.  (This particular expansion has problems with divergences at $z\gtrsim 1$ and will only be used in this paper for comparison with the Riess cosmology.  Instead, the expansion of Linder (2003) will be adopted elsewhere.)  Note that the observed points shift around by a small amount (with an RMS scatter of 0.10 mag about the average difference) as the luminosity relations have to be calibrated for every cosmology.  We see that the bursts with $z>3$ are systematically high, with only 4 out of 18 being below the model curve.  The chi-square for Riess' cosmology is 80.3, which is substantially higher than the chi-square for the concordance cosmology.}

\end{figure}

\clearpage

\begin{figure}
\epsscale{1.1}
\plotone{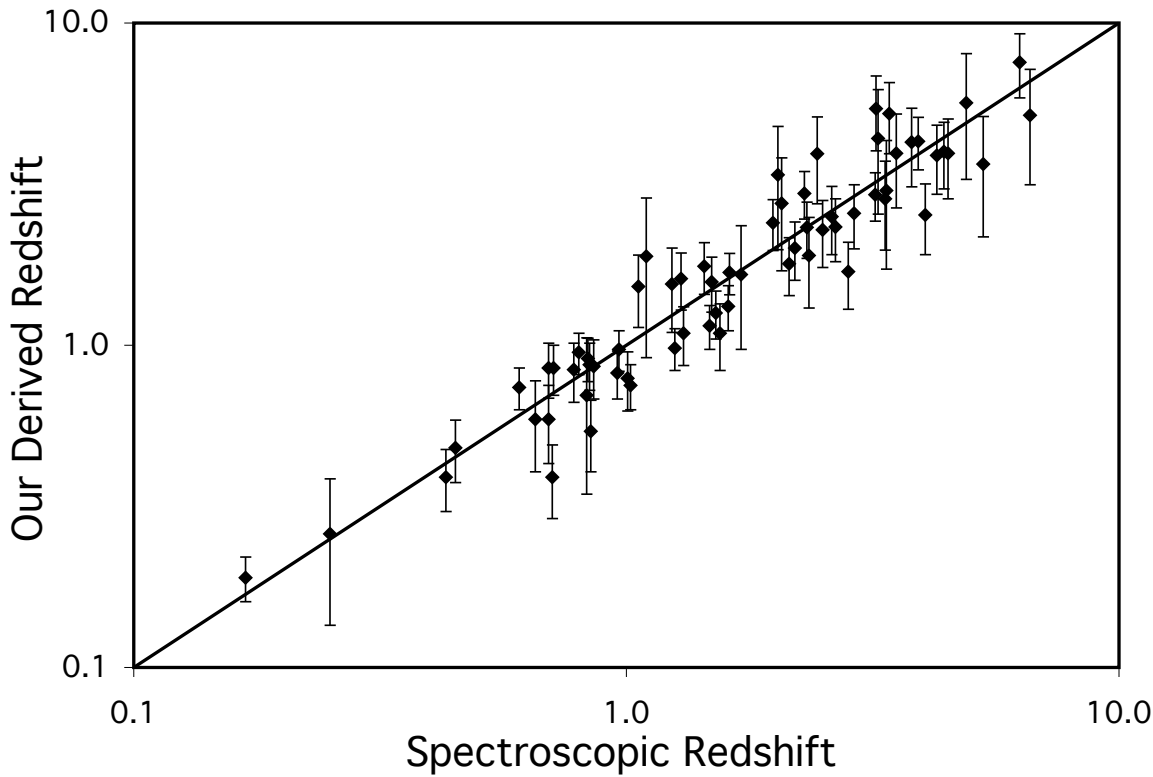}
\caption{
Derived versus known redshifts.  How good are the derived redshifts?  For each burst, the luminosity indicators can be combined to give a distance modulus and then converted to a redshift for some adopted cosmology.  This figure plots the derived versus the known redshifts for all 69 GRBs for the concordance cosmology.  If the derived redshifts were perfect, then they would all be precisely along the diagonal line.  The observed scatter about the diagonal has a reduced chi-square of 1.03, which indicates that the derived error bars are correct.  The scatter about the diagonal does not vary with redshift, which means that the accuracy in the derived values are roughly some constant fraction.  The RMS scatter about the diagonal shows that the average accuracy is 26\%.}

\end{figure}

\clearpage

\begin{figure}
\epsscale{1.1}
\plotone{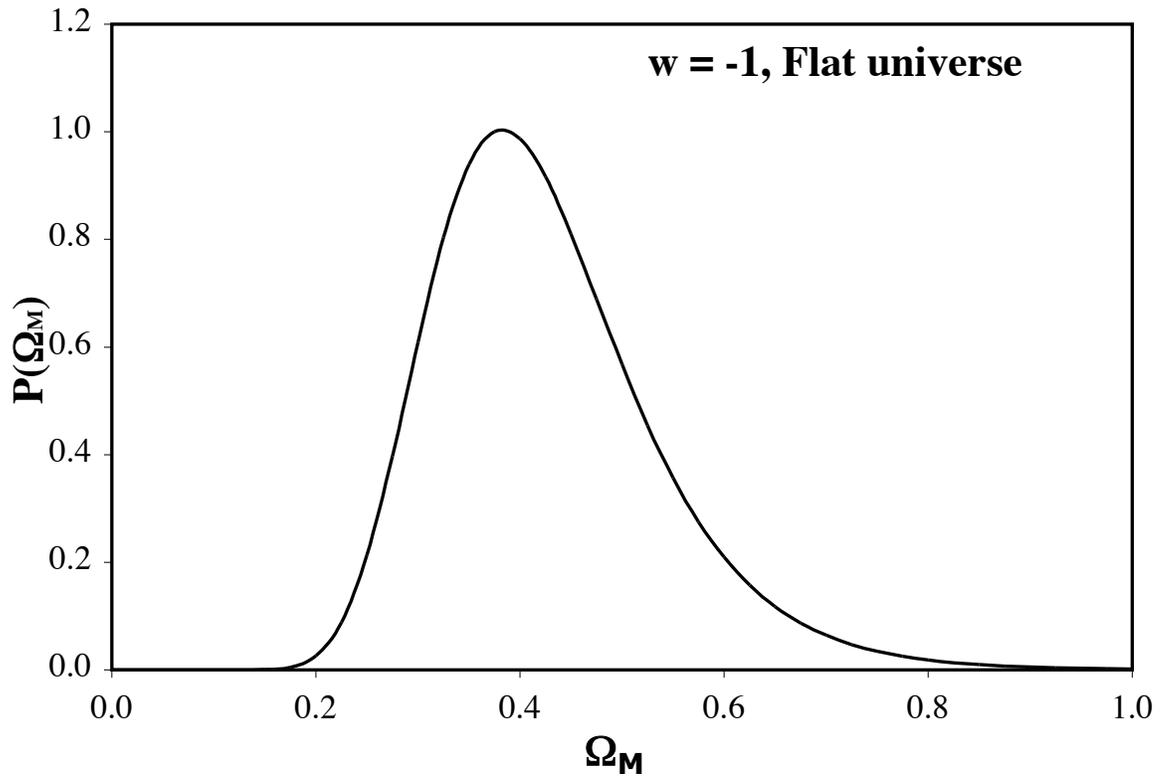}
\caption{
$\Omega_M$ for a flat universe with $w=-1$.  The best value and its asymmetric one-sigma uncertainty is $0.39_{-0.08}^{+0.12}$.}

\end{figure}

\clearpage

\begin{figure}
\epsscale{1.1}
\plotone{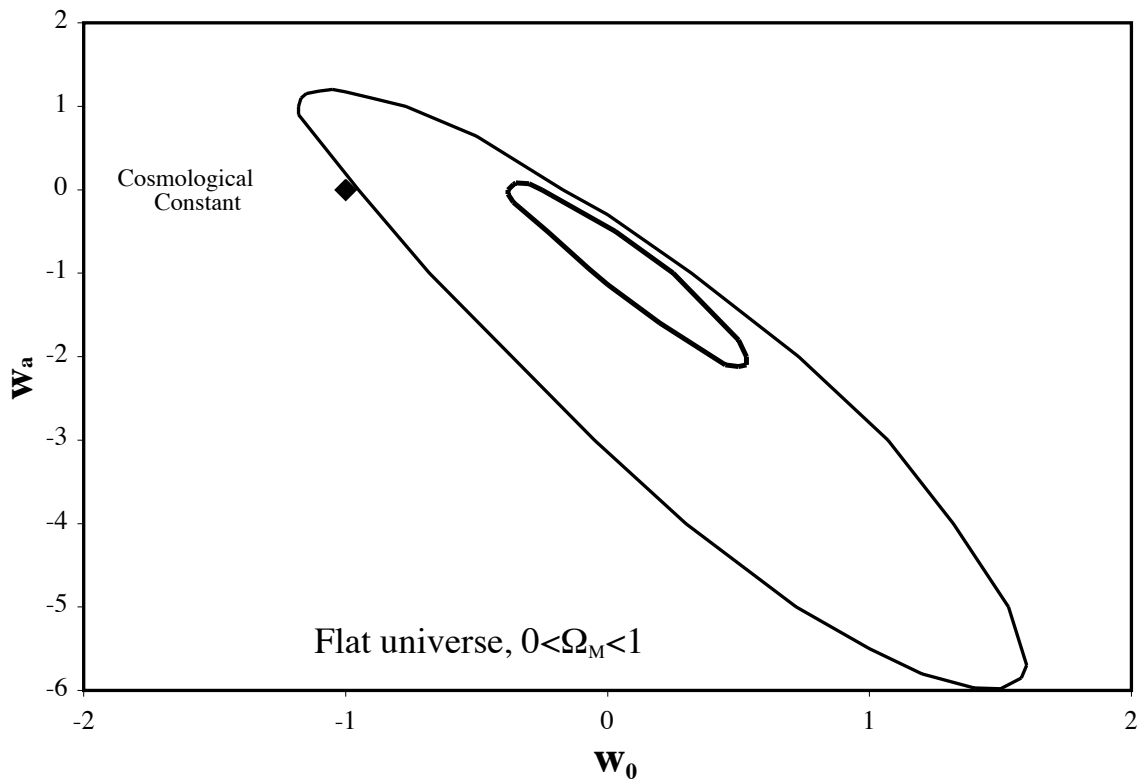}
\caption{
GRB HD results for evolving Dark Energy with a loose constraint on $\Omega_M$.  This plot shows the one-sigma and two-sigma confidence levels for a flat cosmology where $\Omega_M$ was constrained to be between 0 and 1.  The point of best fit implies a relatively shallow slope for $z<0.5$ and this can already be ruled out with supernova data.  Without marginalization, the best fit point has a chi-square of 70.7 for 66 degrees of freedom while the concordance cosmology has a chi-square of 72.3 for 69 degrees of freedom.  This indicates that the concordance model is easily consistent with the GRB HD.}

\end{figure}

\clearpage

\begin{figure}
\epsscale{1.1}
\plotone{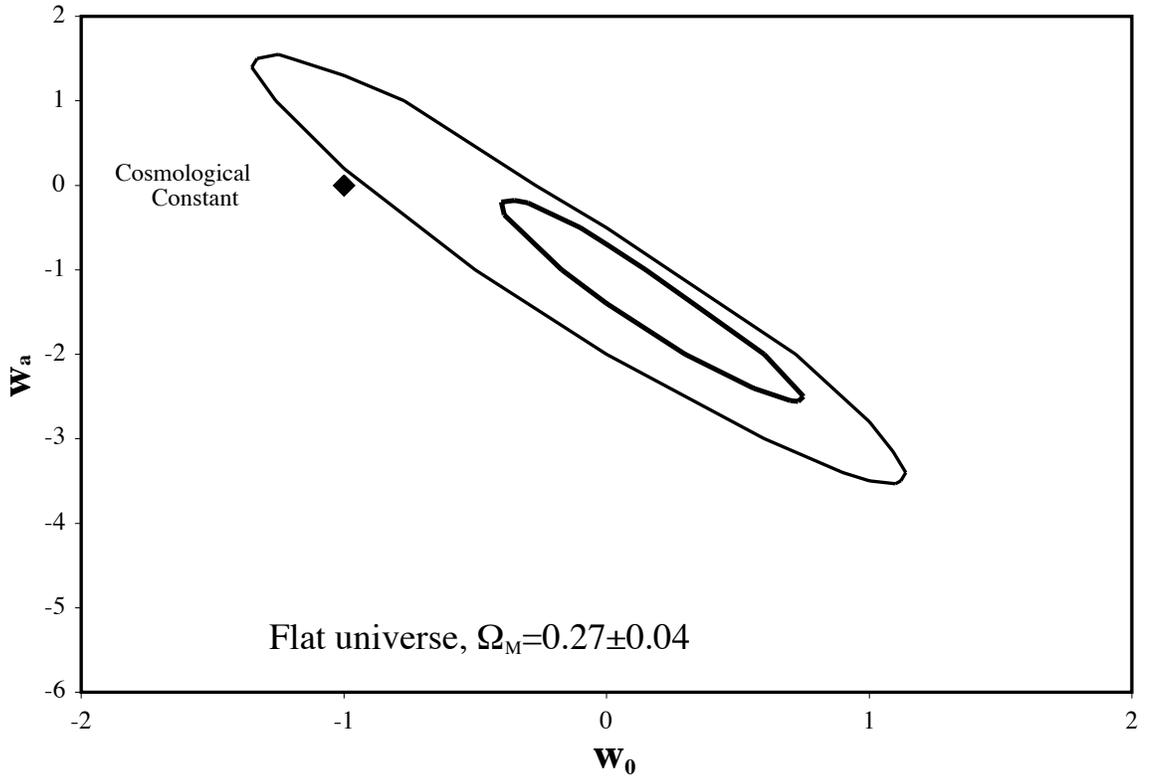}
\caption{
GRB HD results for evolving Dark Energy with a tight constraint on $\Omega_M$.  This plot shows the one-sigma and two-sigma confidence levels for a flat cosmology where $\Omega_M$ was constrained to be $0.27 \pm 0.04$.  The difference between Figures 10 and 11 arises only from a differing constraint on $\Omega_M$.  The same analysis and arguments as in Figure 10 show that a constant Cosmological Constant is acceptable.}

\end{figure}

\clearpage

\begin{figure}
\epsscale{1.1}
\plotone{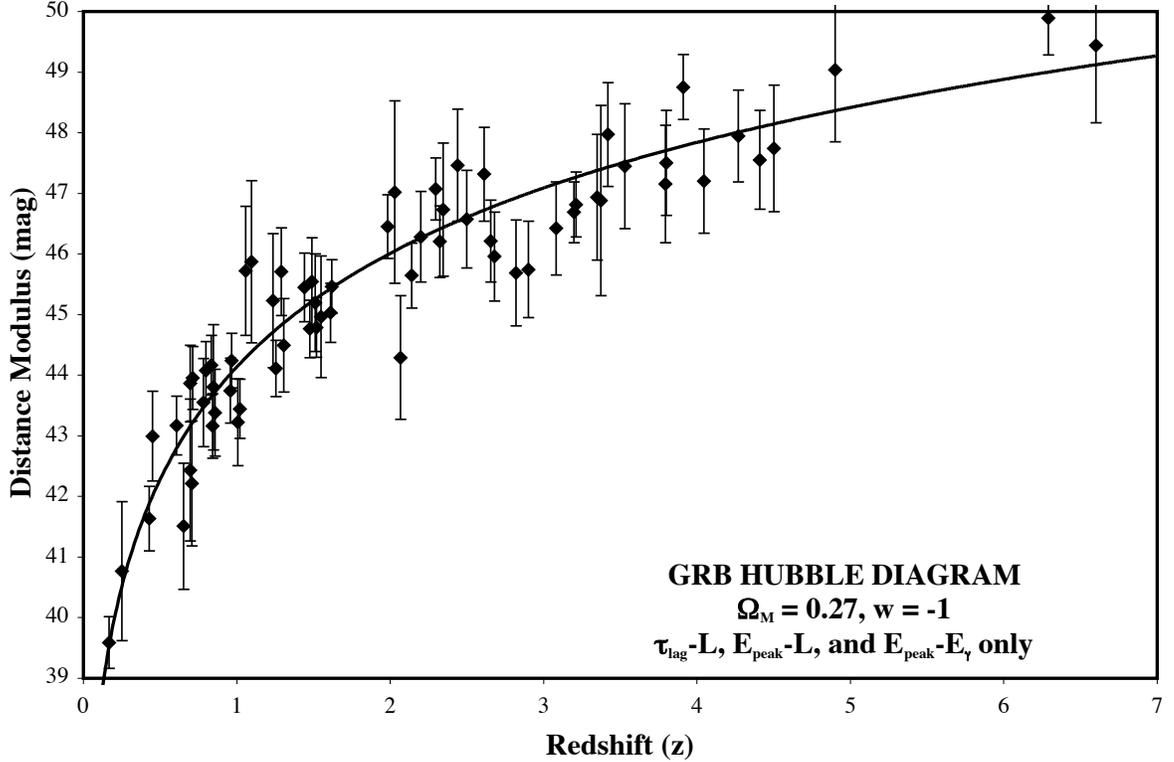}
\caption{
GRB HD with only the $\tau_{lag}-L$, $E_{peak}-L$, and $E_{peak}-E_{\gamma}$ relations.  Do the $V-L$ and $\tau_{RT}-L$ relations add more noise than signal?  This 66 GRB HD was constructed without these two relations and for the concordance cosmology.  This HD shows that the data still follow a smooth curve with no trends.  The scatter about the model curve is larger than that in Figure 7, which is expected due to the $V-L$ and $\tau_{RT}-L$ relations carrying 32\% of the statistical weight.  This demonstrates that the combination of all five luminosity relations is greatly better than using any one or two luminosity relations.}

\end{figure}

\clearpage

\begin{figure}
\epsscale{0.7}
\plotone{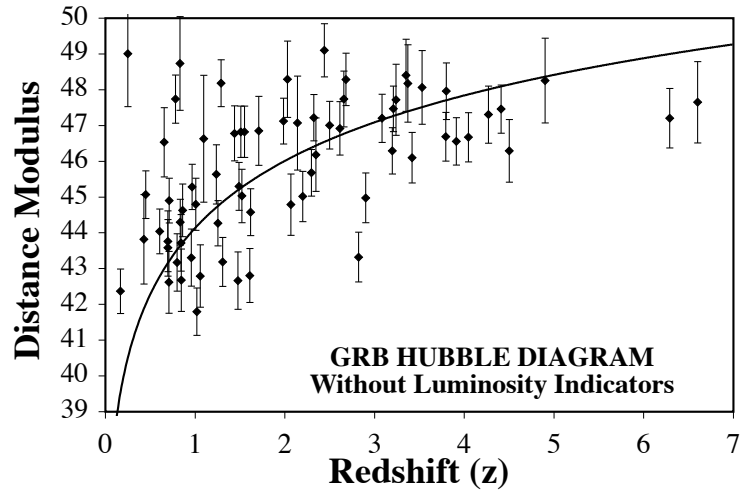}
\caption{
GRB HD with no luminosity indicators.  This small version of Figure 7 displays the HD for the case where the luminosity relations are taken to be flat (i.e., with their returned luminosities being a constant).  This is a blatantly wrong assumption.  The point of this figure is that this HD is horrible with huge scatter, and the comparison with Figure 7 shows that the luminosity relations are producing a vast improvement.}

\end{figure}

\clearpage

\begin{figure}
\epsscale{0.7}
\plotone{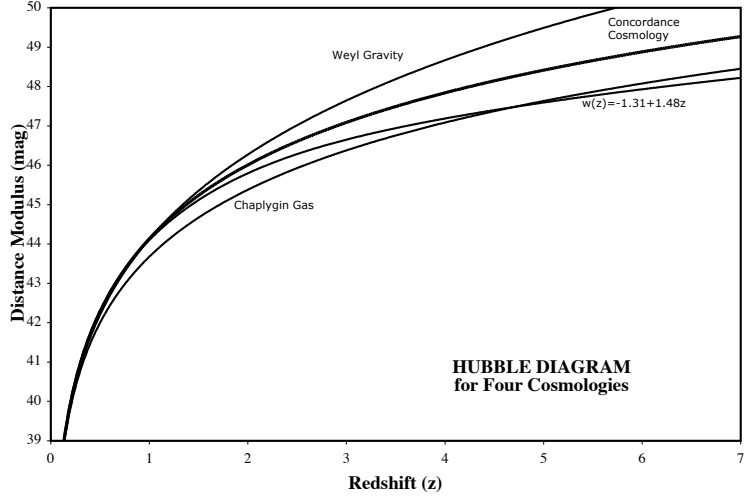}
\caption{
HD for a representative range of cosmological models.  The concordance cosmology (with $\Omega_M=0.27$ in a flat universe with $w=-1$) is now the standard and default based on the supernova measures of the HD up to $z<1$ with a handful up to $z<1.7$.  But many dozens of reasonable models for the equation of state for the Dark Energy as well as for alternatives to General Relativity have been proposed (cf. Szyd\l owski, Kurek, \& Krawiec 2006).  Three represntative alternatives are the Weyl Gravity (Mannheim 2006) with $q_0=-0.2$, a Chaplygin gas (Kamenschik, Moschella, \& Pasquier 2001) with A=0.5, and the 'Riess cosmology' of $w=-1.31+1.48z$ as based on the best fit to the 'gold sample' of supernovae (Riess et al. 2004).  Many of these models show miniscule differences up to $z\sim1.5$ yet large differences for $z>3$.  This figure has three points:  First, many reasonable alternatives to the concordance cosmology exist.  Second, distinguishing between these cosmologies at the relatively low redshifts available to supernovae will be hard due to the small differences that could get hidden by systematic errors (cf. Dom\'{i}nguez, H\"{o}flich, \& Straniero 2001; 2003).  Third, distinguishing between these cosmologies at redshifts $>3$ is easy due to the large differences uniquely measured with GRBs.}

\end{figure}

\end{document}